\documentclass[12pt,preprint]{aastex}

\def\kms{\relax \ifmmode {\,\rm km\,s}^{-1}\else \,km\,s$^{-1}$\fi}

\def\Mso{{$M_{\odot}~$}}
\def\Mj{{$M_J$}}

\def\cm3{${\rm cm}^{-3}~$}

\slugcomment{}

\shorttitle{}
\shortauthors{Villaver, Livio, Mustill \& Siess}

\begin{document}

\title{Hot Jupiters and Cool Stars}

\author{Eva Villaver}
\affil{Universidad Aut\'onoma de Madrid, Department of Theoretical
  Physics, M\'odulo 8, 28049 Madrid, Spain} 
\email{eva.villaver@uam.es}
\author{Mario Livio}
\affil{Space Telescope Science Institute, 3700 San Martin Drive,
Baltimore, MD 21218, USA}
\author{Alexander J. Mustill}
\affil{Universidad Aut\'onoma de Madrid, Department of Theoretical
  Physics, M\'odulo 8, 28049 Madrid, Spain} 
\affil{Department of    Astronomy and Theoretical Physics, Lund
  University, Box 43, SE-221 00 Lund, Sweden} 
\and
\author{Lionel Siess}
\affil{Institut d'Astronomie et d'Astrophysique, 
Universit\'e Libre de Bruxelles, B-1050 Bruxelles, Belgium} 

\begin{abstract}

Close-in planets are in jeopardy as their host stars evolve off the main
sequence to the subgiant and red giant phases. In this paper, we explore
the influences of the stellar mass (in the range 1.5--2\Mso ), mass-loss
prescription, planet mass (from Neptune up to 10 Jupiter masses), and
eccentricity, on the orbital evolution of planets as their parent stars
evolve to become subgiants and Red Giants. We find that planet engulfment
during the Red Giant Branch is not very sensitive to the stellar mass or
mass-loss rates adopted in the calculations, but quite sensitive to the
planetary mass. The range of initial separations for planet engulfment
increases with decreasing mass-loss rates or stellar mass and
increasing planetary masses.
Regarding the planet's orbital eccentricity, we find that as the star
evolves into the red giant phase, stellar tides start to dominate over
planetary tides. As a consequence, a transient population of moderately
eccentric close-in Jovian planets is created, that otherwise would have
been expected to be absent from main sequence stars. We find that very
eccentric and distant planets do not experience much eccentricity decay,
and that planet engulfment is primarily determined by the pericenter
distance and the maximum stellar radius.
 
\end{abstract}

\keywords{Planetary systems; stars: evolution; stars: fundamental parameters; stars: general}

\section{INTRODUCTION}

Planets orbiting evolved stars offer opportunities to explore a wide
range of physical processes that are not applicable to main sequence
hosts. These processes include orbital evolution under the influence
of tides and mass-loss, planetary ejection or evaporation,
instabilities, and the evolution of binary systems
\citep[e.g.,][]{Vl07,Vl09,Dl98,Ds02,Kun11,Bs11,Mv12,Nor10,
Nor13,Mo12,Kp12,Ver11,Ver13}.

Observations show that there is a deficiency of close-in planets
around evolved stars \citep{Joh07b, Lm07, Sat07}, compared to their main
sequence counterparts, even though the two samples have been observed using
the same radial velocity technique \citep[e.g.,][]{Joh07b,Sat08,Wri09,Nied09}. The
masses of these evolved stars derived from stellar model fitting are
typically higher than those of main-sequence planet hosts. At the same
time, evolved stars that are hosting planets do not seem to show differences
in their Galactic velocity distribution from F5-G5 main sequence stars
\citep{Sw13, Mal13} suggesting that the masses of the evolved stars are similar to main-sequence planet hosts.

Two different mechanisms (not necessarily
mutually exclusive) have been proposed to explain the distribution of
planetary orbits around evolved stars. One is based on the dispersal
timescale of protoplanetary disks, and is directly related to the
stellar mass \citep[e.g.,][]{Cur09}, and the other relies on tidal
interaction and orbital decay \citep[e.g.,][]{Vl09}. The former
mechanism requires the masses of the evolved stars to be higher than
those of the main sequence hosts. Whether or not this condition is
consistent with observations has recently been the topic of some
debate \citep{Ll13,Sw13,Joh13}.

The fact remains, however, that both stellar evolution and tidal
interaction still involve considerable uncertainties. For instance,
the prescription used for mass loss affects the orbital evolution both
directly as winds carry away angular momentum, and through its effects on the stellar radius (and thereby
on tidal interaction).  The ratio of planet mass to stellar mass also
strongly affects tides. Previous studies of the orbital evolution of planetary
systems around RGB stars did not investigate either the effects of
the adopted mass-loss prescription or  the planet's eccentricity evolution
\citep[e.g.,][]{Vl09,Kun11}; focused on large orbital distances \citep{Ver11}; or 
on the outcome of the orbital
evolution in a later stage, the Asymptotic Giant Branch phase and its
relevance for the detection of systems during the white dwarf phase \citep{Nor10,Nor13,Mv12}.
Although a range of stellar masses has been
considered by \cite{Kun11} their tidal
evolution calculation was greatly simplified because of the unavailability of the
the internal stellar structure needed for the calculation in their models.

In the present work, we present detailed, self consistent,
stellar models coupled with orbital evolution.  Our
computational grid encompasses the evolution from the main sequence, and up to the onset of Helium
burning in the stellar core. We have selected a range of stellar masses from
1.5 \Mso to 2 \Mso, in steps of 0.1 \Mso, and for each stellar mass we have
computed the evolution of the star with 3 different prescriptions for
the mass-loss. For each star, we have calculated the orbital evolution
with planetary masses corresponding that of Neptune, and to 1, 2, 5,
and 10 Jupiter masses.  We have also considered the 
planet's eccentricity evolution. Our goal is to characterize and
quantify the effects of mass-loss, stellar mass (in an interesting
range),  planetary mass and eccentricity on planet survival along the Red Giant
Branch (RGB) phase.

\section{THE CALCULATIONS}
We follow the same procedure as in \cite{Vl09} (from now on
VL09) and \cite{Mv12} to determine the planet orbital and
eccentricity evolution as the star evolves off the Main Sequence(MS)
and up the RGB. We take into account the changes in the 
mass of the star, the gravitational and frictional drag, and the tidal force. 
We have not evaluated in this paper either the accretion onto, or the
ablation of matter from, the surface of the planet. Based on
the estimates provided by the prescription
given in VL09 we find that the effect of  mass-loss from the star dominates
over those processes, even when small RGB mass-loss
rates are considered.

We use three different prescriptions
for the mass-loss in this paper; two that follow a Reimers' law
valid  for Red Giants \citep{Rei75} and use  values of $\eta_R$  (the Reimers
parameter) of  0.2 and 0.5 in the formula,
\begin{equation}
\dot{M}_\mathrm{R} = 4 \times 10^{-13}\, \eta_R \frac{L_\star
  R_\star}{M_\star} \quad [M_{\sun}~\mathrm{yr}^{-1}]~~,
\label{Rei}
\end{equation}
and a third one based on
the new semiempirical relation provided by \cite{Sch05}
\begin{equation}
\dot{M}_\mathrm{Sc} =  \eta \frac{L_\star R_\star}{M_\star}
\left( \frac{Teff}{4000 K}\right) ^{3.5} \left( 1+\frac{g_{\sun}}{4300
    g_\star}\right) \quad [M_{\sun}~\mathrm{yr}^{-1}]~~,
\label{Sch}
\end{equation}
where $L_\star$, $R_\star$, $g_\star$ and $M_\star$ are the stellar luminosity, radius,
surface gravity, and
mass respectively (in solar units) and the factor $\eta = 8 \times 10^{-14}$ has been obtained by fitting
the theoretical relation to observations of globular clusters with
different metallicities;
$g_{\sun}$ is the Sun's surface gravity. Thus, the rate of change of the stellar mass is simply 
given by $\dot{M_\star}= - \dot{M}_\mathrm{R}$, or  $\dot{M_\star}=
-\dot{M}_\mathrm{Sc} $. The stellar evolution calculations have been
self-consistently carried out using these 3 mass-loss prescriptions
(see \S 2.1 for details).

We consider that a planet of mass $M_p$ and radius
$R_p$ is orbiting with a velocity, $v$ a star of mass $M_\star$. 
Conservation of angular momentum gives the equation for the rate of
change in the semimajor axis of the planet assuming that the
time-scale for mass loss is much greater than the orbital time-scale (see, e.g.,
\citealt{Ale76,Ls84, Vl09,Mv12}), 
\begin{equation} 
\left(\frac{\dot{a}}{a}\right) =-\frac{\dot{M_\star}}{M_\star+M_p}-\frac{2}{M_p v}
\left[F_f+F_g\right]-\left(\frac{\dot{a}}{a}\right)_{t}~~,
\label{todo}
\end{equation}
where $({\dot{a}}/{a})_{t}$ is the rate of orbital decay due to the
tidal interaction and $F_f$ and $F_g$ are respectively the frictional and gravitational drag
forces which have been computed with the prescription given in VL09.
The frictional force is expressed in the form \citep[e.g.,][]{Rosen63}  
\begin{equation}
F_f = \frac{1}{2} C_d \rho v^2 \pi R_p^2~~,
\label{ff}
\end{equation}
where $C_d\simeq0.9$ is the dimensionless drag coefficient for a
sphere, and the gravitational drag force, $F_g$ has the functional form given by \citep[e.g.,][and
references therein]{Ost99} 
\begin{equation} 
F_g= 4\pi \frac{(G M_p)^2 }{c_s^2}\rho I~~,
\label{fg}
\end{equation}
where for $I$ we used the value $I\simeq0.5$ which is appropriate for
the range of Mach numbers for this problem.  

We consider tides acting on both the planet and the star. For the planetary
tides, we use the formalism of \cite{Dd04} with a fixed tidal
dissipation efficiency parameter
$Q_\mathrm{pl}^\prime$, while for the stellar tides, we adopt the
formalism of \cite{Zah77}, in which tidal energy is dissipated by
turbulent motions in the star's convective envelope. 
In fact, in giant stars, which have massive convective envelopes, the
most efficient mechanism to produce tidal friction is turbulent viscosity
\citep[e.g.,][]{Zah66,Zah77,Zah89}.  For the angular momentum loss associated to the tidal term
$({\dot{a}}/{a})_{t}$, the dissipation timescale is determined by the
effective eddy viscosity, with eddy velocities and length scales given
approximately by standard mixing length theory if convection
transports 
most of the energy flux \citep{Zah89,Vp95,Retal96}. 
The stellar tidal term is given by
\begin{equation}
\left(\frac{\dot{a}}{a}\right)_{t} = \frac{1}{9\tau_d} \frac{M_\mathrm{env}}{M_\star} q (1+q)
\left(\frac{R_\star}{a}\right)^8\times
\left[2f_2+e^2\left(\frac{7}{8}f_1-10f_2+\frac{441}{8}f_3\right)\right], 
\label{eq:adotstar}
\end{equation}
and the planet eccentricity decays as
\begin{equation}
\frac{\dot
  e}{e}=  -\frac{1}{36\tau_d} \frac{M_\mathrm{env}}
{M_\star} q (1+q) \left(\frac{R_\star}{a}\right)^8 \times
\left[\frac{5}{4}f_1-2f_2+\frac{147}{4}f_3\right],
\label{tidal}
\end{equation}
with $M_\mathrm{env}$ and $R_\mathrm{env}$ being the mass and radial extent of the  convective
envelope respectively, $q =
M_p/M_\star$, and $\tau_d$ is the eddy turnover timescale in the stellar
envelope 
 \begin{equation}
\tau_d=\left[\frac{M_\mathrm{env}\left(R_\star-R_\mathrm{env}\right)^2}{3L_\star}\right]^{1/3}.
\end{equation}
The frequency components $f_i$ are given by
\begin{equation}
f_i=f^\prime\min\left[1,\left(\frac{2\pi}{inc_\mathrm{F}\tau_d}\right)^\gamma\right],
\label{gamaa}
\end{equation}
where $in$, with $n$ the mean motion, are the individual frequency
components. As in VL09 and \cite{Mv12} we have used $c_\mathrm{F}=1$,
$f^\prime=9/2$ and a value of $\gamma =2$ which is consistent with the
results obtained from numerical calculations \citep[e.g.,][]{Zah77,Gn77, Pen07}.

In this study, we have made a particular choice, turbulent convection,
for the dissipation of the equilibrium tide. So under our assumption,
if the star does not have a convective envelope, there is no
tide. For stars with radiative envelopes, the use of the $Q$ formalism circumvents this problem by
parameterizing the strength of the tidal forces into a variable
$Q'_{\star}$. When  the $Q'_{\star}$ parameter is calibrated  using MS stars
has values in the
$10^5-10^{10}$ range \citep[e.g.,][]{Jac08a,Jac09,Pen11}.  
However, for giant stars, a much smaller 
$Q'_{\star} \approx 10^2-10^{3}$ is found by
\cite{Nor10} to be equivalent  to the \cite{Zah77}
formalism. Note that since giant stars have deep convective zones the
most appropiate mechanism to describe the equilibrium tide is the one
adopted here \citep[e.g.,][]{Ogil14}. Thus it is inappropriate to apply
the $Q$ formalism using the MS calibration for our
stars and will anyhow result on much weaker tides than
assumed.  We have neglected other
tidal mechanisms such as dynamical tides \citep[e.g.,][]{Ws02}, which
although may be important for massive stars during the MS
they are mostly irrelevant for the computations for convective stars presented here.   

Only during the MS phase we might have a non-convective star. Thus for stars without a convection
zone and just in order to avoid numerical round-off errors 
we set the convective time-scale to an arbitrarily large value ($10^9$\,years).  Note that during the MS
we are only studying the eccentricity evolution for which only
the planetary tides are important.

For eccentric orbits we have also implemented a planetary tide using
the standard $Q$ model \citep[e.g.,][]{Matsu10}. The planet's semi-major axis, eccentricity and spin rate then evolve
according to, 
\begin{eqnarray}
\frac{\dot
  e}{e}&=&-\frac{81n}{2Q_\mathrm{p}^\prime}\frac{1}{q}\left(\frac{R_\mathrm{p}}{a}\right)^5 \times
\left[g_3\left(1-e^2\right)^{-13/2}-\frac{11g_4\left(1-e^2\right)^{-5}\Omega_\mathrm{p}}{18n}\right]\label{eq:edotpl}\\
\frac{\dot a}{a}&=&\frac{2e\dot e}{1-e^2}-\frac{9}{Q_\mathrm{p}^\prime}\frac{1}{q}\left(\frac{R_\mathrm{p}}{a}\right)^5\times\left[g_2\left(1-e^2\right)^{-13/2}n-g_5\left(1-e^2\right)^{-5}\Omega_\mathrm{p}\right]\label{eq:adotpl}\\
\dot\Omega_\mathrm{p}&=&\frac{9n^2}{2\alpha_\mathrm{p}Q_\mathrm{p}^\prime}\frac{1}{q}\left(\frac{R_\mathrm{p}}{a}\right)^3
\times \left[g_2\left(1-e^2\right)^{-6}-g_5\left(1-e^2\right)^{-4.5}\frac{\Omega_\mathrm{p}}{n}\right].\label{eq:odotpl}
\end{eqnarray}
where $R_\mathrm{p}$ is the planet's radius.
The eccentricity functions $g_i$ are given by
\begin{eqnarray}
g_2&=&1+\frac{15}{2}e^2+\frac{45}{8}e^4+\frac{5}{16}e^6\\
g_3&=&1+\frac{15}{4}e^2+\frac{15}{8}e^4+\frac{5}{64}e^6\\
g_4&=&1+\frac{3}{2}e^2+\frac{1}{8}e^4\\
g_5&=&1+3e^2+\frac{3}{8}e^4 .
\label{gs}
\end{eqnarray}

We note that, while the planet tidal equations are valid for arbitrary
planet eccentricities, the stellar tidal equations are based on the
lowest-order expansion of the equations and are not strictly
valid when the planetary eccentricity is high (note that in
\cite{Zah89} the $\dot a$ equation is valid for any eccentricity
but he uses the weak approximation). The reason for this lies in
the stellar tidal model; the different Fourier components have
different frequency dependencies, and as the eccentricity rises,
higher-frequency Fourier components come into play. Nevertheless, for
this exploratory study, we used the truncated equations even for highly
eccentric planets, deferring a full expansion of the tidal forces to
future work. It is important to keep in mind as well that, as
discussed in \cite{Mv12}, the parameters in the tidal 
equations are poorly known, and consequently the tidal forces may be
stronger or weaker than assumed here. 

The main uncertainty in the tidal model used comes from the
need to reduce the effective viscosity when the tidal
period is short compared to the typical convective timescale
\citep[e.g.,][]{Ogil14}. This is taken into account by variations in the power index
$\gamma$ in Eq. \ref{gamaa}. As $\gamma$ increases the tidal
disipation becomes less efficient but only when the planet's  orbital period is shorter than the eddy turnover
timescale in the stellar envelope. We have chosen a $\gamma =2$ in
Eq. \ref{gamaa} consistent with the results of \cite{Ogil12}, earlier results of \cite{Pen07} suggested a $\gamma =1$.

The tidal model is only sensitive to the assumed value of $\gamma$ for
small orbital distances when the stellar radius is small
\citep[e.g.,][]{Kun11}. Thus variations in the power index will not lead
to any significant changes in the final outcome of our planets 
since  we are
dealing with stars that reach a large radius at the tip of the RGB.  If the planet has such a short orbital period 
to be sensitive to the adopted $\gamma$ then it would be engulfed very early by the expansion of the star.

Measurements of the spin rates of giants provide average
values of $ v sin i  \leq 2$\kms
\citep[e.g.,][]{Deme96,Mass08}, with rapid rotators found only in a
few percent (1-2 \%) of giant stars \citep{Car11}. 
These observations justify our assumption of using
non-rotating stars for the calculation of the tidal
forces. Planet spin rates were allowed with values between 10 and 1000
radians per year.

A proper calculation of the stellar tides requires knowledge of the stellar structure (i.e. $M_\mathrm{env}$ and
$R_\mathrm{env}$). Note that some calculations in the
literature, do not have this information and as result they rely on
simplifications such as the assumption that $M_\mathrm{env}$ = $M_\star$ and
$R_\mathrm{env}$ = $0$ for the evaluation of the effects of  tidal dissipation
on the planet's orbital evolution.

\subsection{The Stellar Models}

The stellar evolution models were calculated with the STAREVOL code
\citep{Siess06}. We have computed a small grid of non-rotating models with
initial masses in the range 1.5 and 2 \Mso in mass steps of 0.1 \Mso. 
Based on the observations by \cite{Mal13} we 
have chosen a stellar metallicity that fullfils  two requirements: 
i) it is consistent with the observed metallicity of planet-hosting giant stars
with stellar masses M$_\star>$ 1.5 \Mso; and ii) it agrees with the observed metallicities 
of subgiant stars with planets detected.  The
initial metallicity is thus set to [Fe/H]=0.19 with a composition scaled solar
according to \cite{GNS96}.
We do not consider any ``extra-mixing processes''
such as overshooting or thermohaline mixing and use the Schwarzschild
criterion to define the convective boundaries. We adopt
$\alpha_{\mathrm{MLT}}=1.75$ for the mixing length parameter which value
was determined from solar fitting models.  As described above, we
considered three representative mass-loss rate prescriptions (see Eqs. \ref{Rei} and \ref{Sch}).

The extent of the RGB on the HR diagram is very sensitive to the stellar
mass for values around the transition mass that marks the border
between degenerate and non-degenerate cores. We use here RGB models with masses lower than 2.0 \Mso
because they develop electron degenerate He cores after the end of
central H-burning for Solar metallicity. These stars have an extended and luminous
RGB phase prior to Helium ignition and therefore represent the
most interesting arena for the problem analyzed in this paper. The
maximum stellar luminosity at the tip of the RGB is reached for stars
with degenerate
cores and it translates into a maximum stellar radius reached during the RGB. 
Higher mass stars can ignite
helium quietly, terminating the ascent up the RGB before electron
degeneracy becomes appreciable in the core, and they reach smaller
radii than stars with degenerate cores.  Note, however, that the precise value of the transition mass
between low and
intermediate mass stars depends on the initial chemical composition \citep[e.g.,][]{Swe89,Swe90}.
Decreasing the  initial Helium abundance or increasing the heavy
element abundances leads to higher transition stellar masses.
Thus for stars born in higher metallicity environments,  we have
larger initial masses marking the transition between degenerate and
non-degenerate  He cores. 

We summarize in Table~1 some of the properties of our stellar models. 
Column (1) gives the initial mass; column (2) the 
mass-loss prescription used where Sc refers to the \cite{Sch05} model and $\eta_R=0.5+OS$ is a model with overshooting; column (3) gives the stellar mass at the time of the deepest extent of the convective envelope during the first Dredge-up (1DUP); columns (4), (5), and (6) give the stellar mass, radius and luminosity at the tip of the RGB respectively; column (7) provides the mass of the Hydrogen depleted core; column (8) the mass-loss rate at the tip of the RGB, and column (9) gives  the duration of the RGB phase between the time of the 1DUP and the RGB tip. Finally column (10) gives the minimum initial orbital distance to avoid engulfment at the tip of the RGB calculated for a planet with the mass of Jupiter.

\section{RESULTS}
As hydrogen becomes exhausted at the center of the star,  core contraction
accelerates and the star leaves the main sequence on the HR diagram. 
The major energy production source shifts to a thick shell outside the core where shell ignition
drives envelope expansion and causes the stellar radius to increase
\citep[e.g.,][]{Iben67}. Shortly after the star reaches the base of the RGB, convection
expands inward from the surface and the first dredge-up takes place.

The planet's orbital evolution was computed 
by solving the system of Eqs. \ref{todo} -- \ref{gs} coupled with the stellar structure evolution. 
We did so for each of the six stellar masses
and three mass-loss prescriptions considered (a total of 18 stellar
models)  and we used five planetary masses. For each one of these 90 models we then calculated the orbital evolution
by varying the initial orbital separation between 0.1 and
3.5 $\mathrm{AU}$ using steps of 0.01 $\mathrm{AU}$. 
Every integration timestep included an update of the stellar structure
(fundamental for an accurate calculation of the stellar tide)
and of the variable mass-loss rates. We
have introduced also a constraint in the orbital
integration timestep so that it was never larger than the timestep at which a
significant change in the stellar structure took place.

We first consider circular orbits ($e = 0$) in Eqs. \ref{eq:adotstar}
to \ref{gs} to investigate the effects of changing
the stellar models.
For the systems considered in this paper, tidal dissipation in the
star dominates (see e.g. \citealt{Matsu10}), and thus we justify circular orbits (for now) on the basis
of the fact that both the eccentricity and the semi-major axis damp on similar time-scales
\citep{Mv12}. Note as well that the initial value of
the eccentricity has little effect on the orbital decay rate
\citep[e.g.,][]{Jac08a,Jac08b}.  

We first focus on estimating variations induced by 
varying parameters such as mass-loss, stellar mass or planet mass, and we defer the discussion of the eccentric orbits to \S 3.4. 

The combined effects of tidally induced orbital decay and mass-loss
induced orbital expansion modifies the orbit of the planets in
a simple way. If the initial planet separation is within a
certain range of distances from the star, the planet will
experience an orbital decay caused by the tidal
interaction. This ends up with
the planet plunging into the stellar envelope. If, on the other hand,
the initial orbit is beyond a certain radius, the planet avoids engulfment.

\begin{figure}
\epsscale{1.10}
\plotone{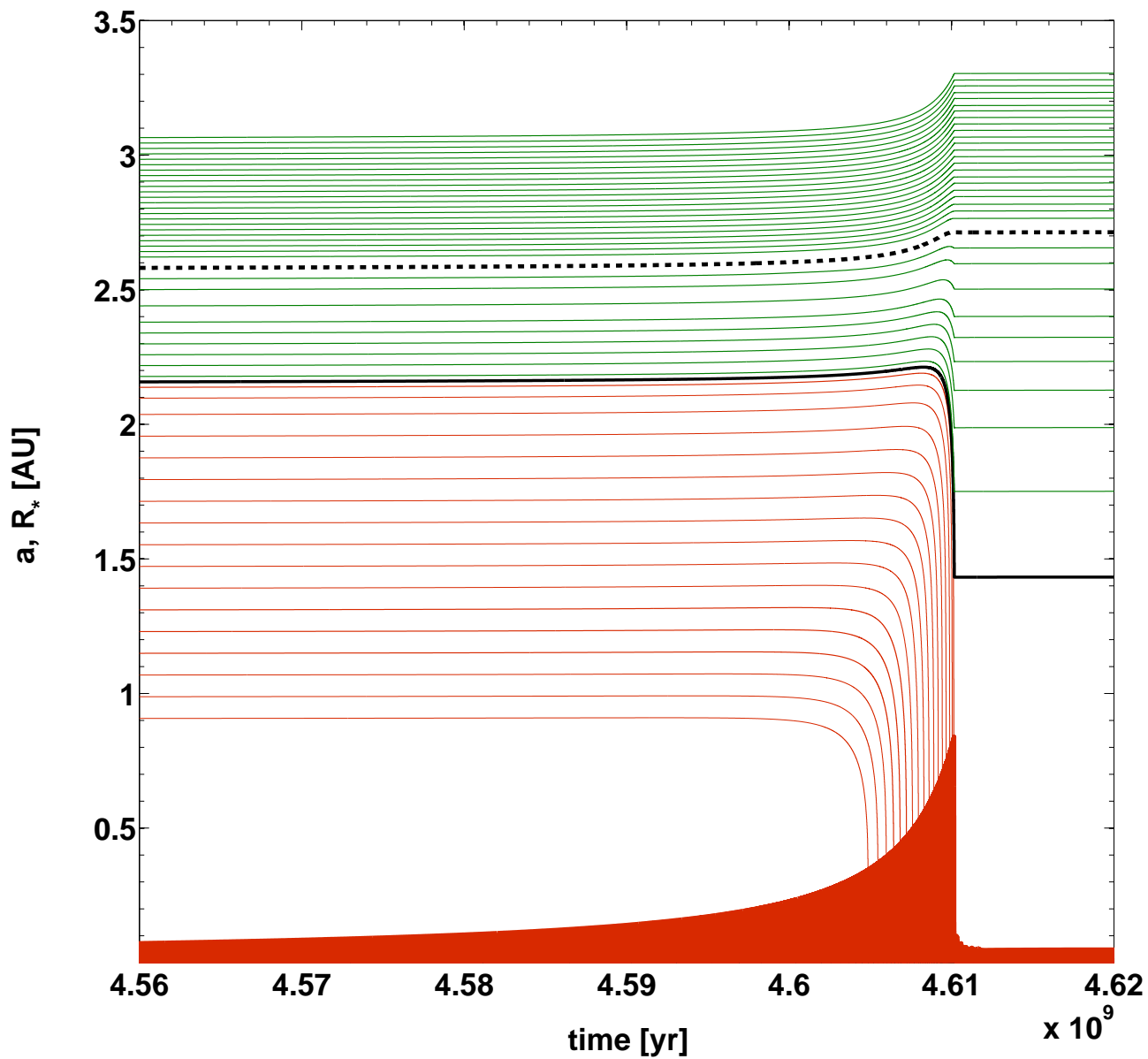}
\caption[ ]{The filled red area represents 0.06 Gyr in the evolution of the
  stellar radius as the star goes to the tip of the RGB. The star has a mass
  of 1.5 \Mso  and is evolving under the Schr{\"o}der 
\& Cuntz(2005) mass-loss prescription. The evolution of a set of 
orbits of a planet with 1 Jupiter mass is shown as well, with the
red lines representing a set of initial separations for which
the planet ends up inside the stellar envelope. The green lines represent
the initial separations for which the planet avoids engulfment. The solid black line is the minimum
initial orbit for which the planet avoids being engulfed and the dashed black line marks the initial orbit beyond which
the planet is not affected by the tidal forces. 
\label{Fig1_evo}}
\end{figure}

Fig.~\ref{Fig1_evo} shows  Jupiter mass planet's orbits 
during $\approx$ 0.06 Gyr along the RGB phase  for a range of initial orbital distances and a
particular stellar model (1.5 \Mso with the \citealt{Sch05}
mass-loss prescription). A few features in
Fig.~\ref{Fig1_evo} are worth noting. First, marked with a solid black line is the initial orbital distance beyond
which the planet avoids falling into the stellar envelope during the
RGB phase. Every initial separation below the solid black line terminates with the planet being engulfed. 

What happens to those planets that enter the envelope largely depends
on the ratio of the planet's mass to the envelope mass. Many will end up
merging with the stellar core, given the fast decay of the
orbit induced by the strong drag forces  \citep{Nor06,SL99a,SL99b}. Rough estimates
of planet survival inside stellar envelopes provide minimum masses
of the order of 10-15 Jovian masses (again, depending also on the
envelope's mass; \citealt{Vl07, Nor10}). An analysis of
the possible progenitors of the planets found orbiting the
horizontal branch star KIC 05807616 \citep{Cha11} also concludes that 
the surviving planets likely had a mass of a few Jupiter masses
\citep{Pas12}. Note that for
survival to occur, a planet must be able
to supply enough of its orbital energy to the stellar envelope to
unbind the latter before the planet spirals into the disruption region or the
stellar core. Unfortunately, the efficiency of the process of
unbinding (commonly parameterized with $\alpha_{CE}$,
e.g. \citealt{Ls88}) is rather uncertain. Given that the largest planet mass we are considering
is 10 \Mj, our working hypothesis in this paper is that whenever the planet
gets inside the stellar envelope it would be destroyed. We defer
further discussion of planetary systems entering stellar envelopes for a
forthcoming study.

The second important feature of the planet's orbital evolution 
that is important to mention is the range of
initial orbital distances that experience tidal decay but that still manage to
avoid entering the surface of the star. This range of orbits lies
in Fig.~\ref{Fig1_evo}  close to the solid line. The 
dot-dashed black line shows a typical example of an orbit dominated by
mass-loss. The orbital decay
can be substantial for initial orbits close to the critical limit
(marked by the solid black line) with the planet ending up in a
significantly tighter orbit than the initial one. Planets that
start at orbital distances sligthly larger than marked by the solid
line, still experience the consequences of tidal forces,
but once helium burning is ignited in the core, the star
contracts and the interaction stops. Planets in this initial range do not reach the stellar
envelope but end up at smaller separations.

Beyond the dot-dashed line in Fig.~\ref{Fig1_evo} the orbits increase
due to mass-loss from the system. 

\subsection{The Effects of RGB Mass-Loss}

Eq.~\ref{todo} clearly suggests that the critical orbital distance
for engulfment, and the distance beyond which the separation is expected to
increase are sensitively dependent on the mass-loss prescription adopted in the
calculations (for an analytical treatment of the problem see
\citealt{Ada13}). The purpose of our self-consistent calculation, with models under 
different mass-loss prescriptions, is to quantify the influence 
of this relatively poorly constrained parameter on 
planet survival during the RGB evolution. 

\begin{figure}
\epsscale{1.10}
\plotone{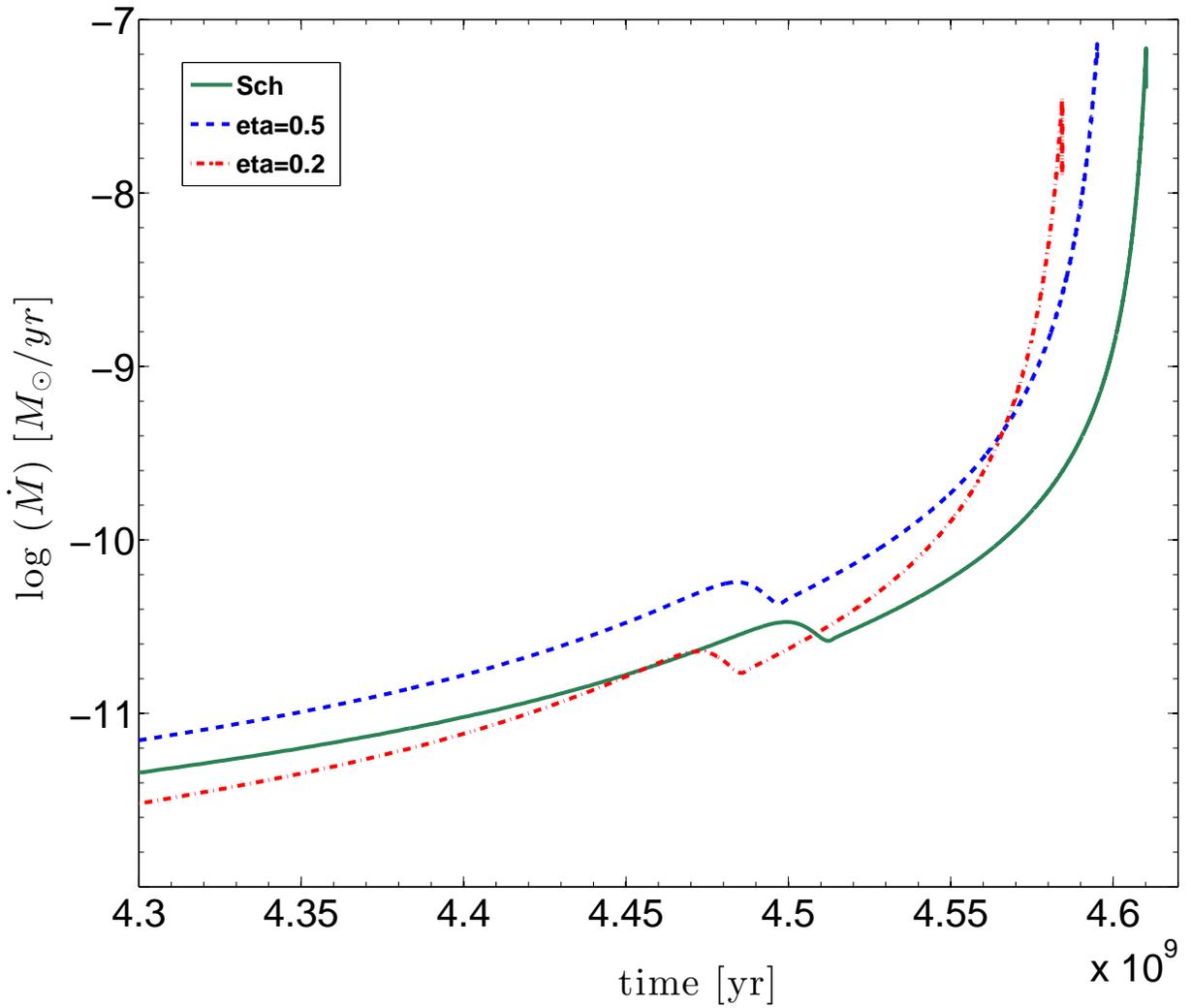}
\caption[ ]{Mass-loss rates for 1.5 \Mso
stars during the last years of the RGB evolution under $\eta_R =0.2$
(red dot-dashed), $\eta_R =0.5$ (blue dashed), and Schr{\"o}der  
\& Cuntz (2005) (green solid line). 
\label{massloss}}
\end{figure}

\begin{figure}
\epsscale{1.10}
\plotone{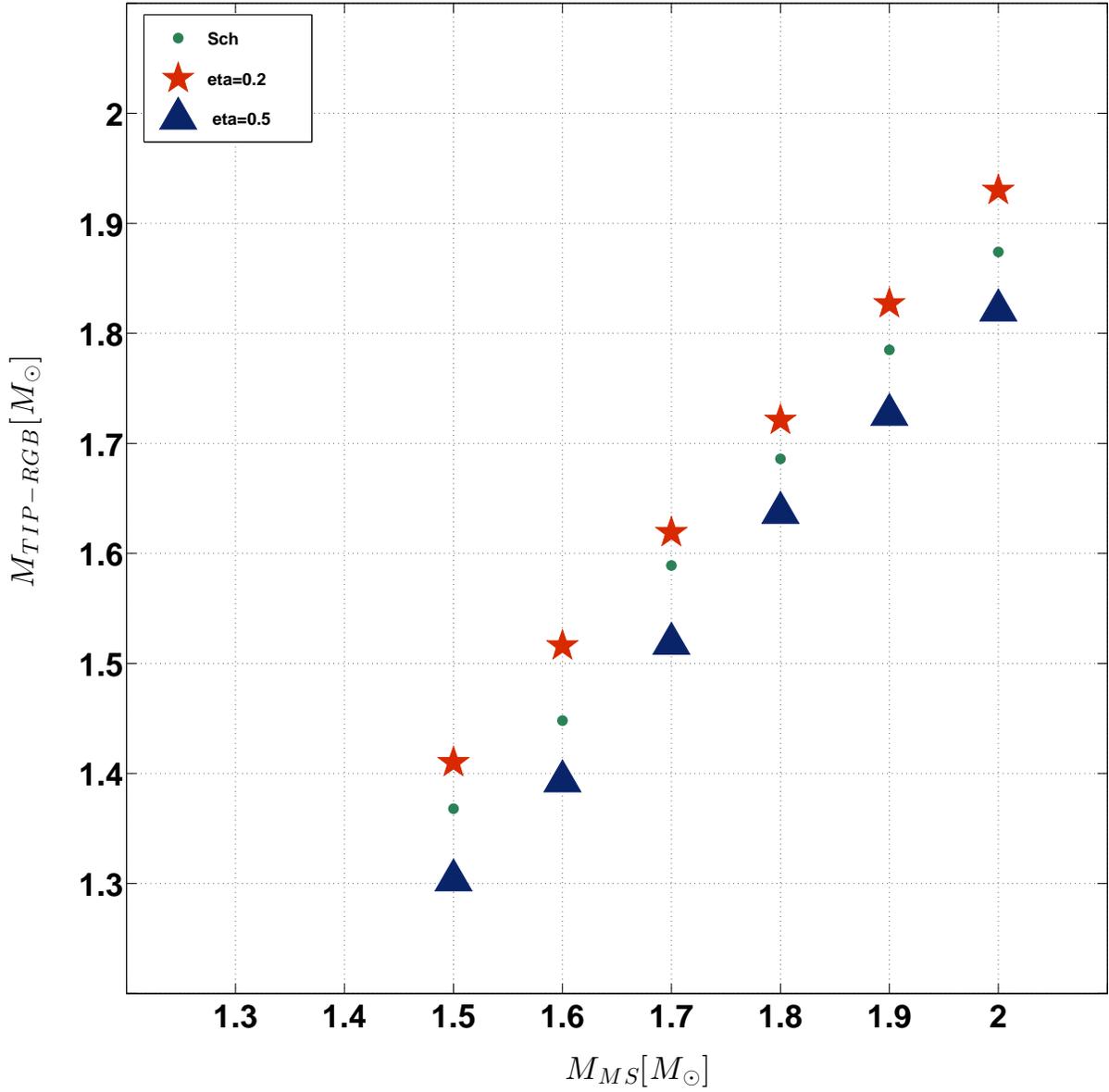}
\caption[ ]{Mass of the star at the tip of the RGB in solar units
  versus the MS mass of the star. The different symbols correspond to
  the 3 prescriptions used and are indicated in the legend  (red stars
  ($\eta_R =0.2$), blue triangles ($\eta_R =0.5$), and green asterisks(Schr{\"o}der 
\& Cuntz 2005). 
\label{mimf}}
\end{figure}

Along the RGB the models under different
mass-loss prescriptions show significant differences associated
entirely with the way these stars are losing mass (see Table 1).  The mass-loss rate during the 
RGB is relatively smooth (see Eqs \ref{Rei} and
\ref{Sch}) and never reaches high values (see Fig.~\ref{massloss}).
The largest and smallest mass-loss rates are attained for the $\eta_R
=0.5$, and $\eta_R =0.2$ prescriptions respectively, with the
\cite{Sch05} model having mass-loss rates in between these two values.
In Fig.~\ref{massloss} we have plotted the evolution of the mass-loss under
the 3 prescriptions used for the 
1.5 \Mso model. In Fig.~\ref{mimf} we show 
the stellar mass reached at the tip of the RGB versus the stellar MS mass. 

\begin{figure}
\epsscale{1.10}
\plotone{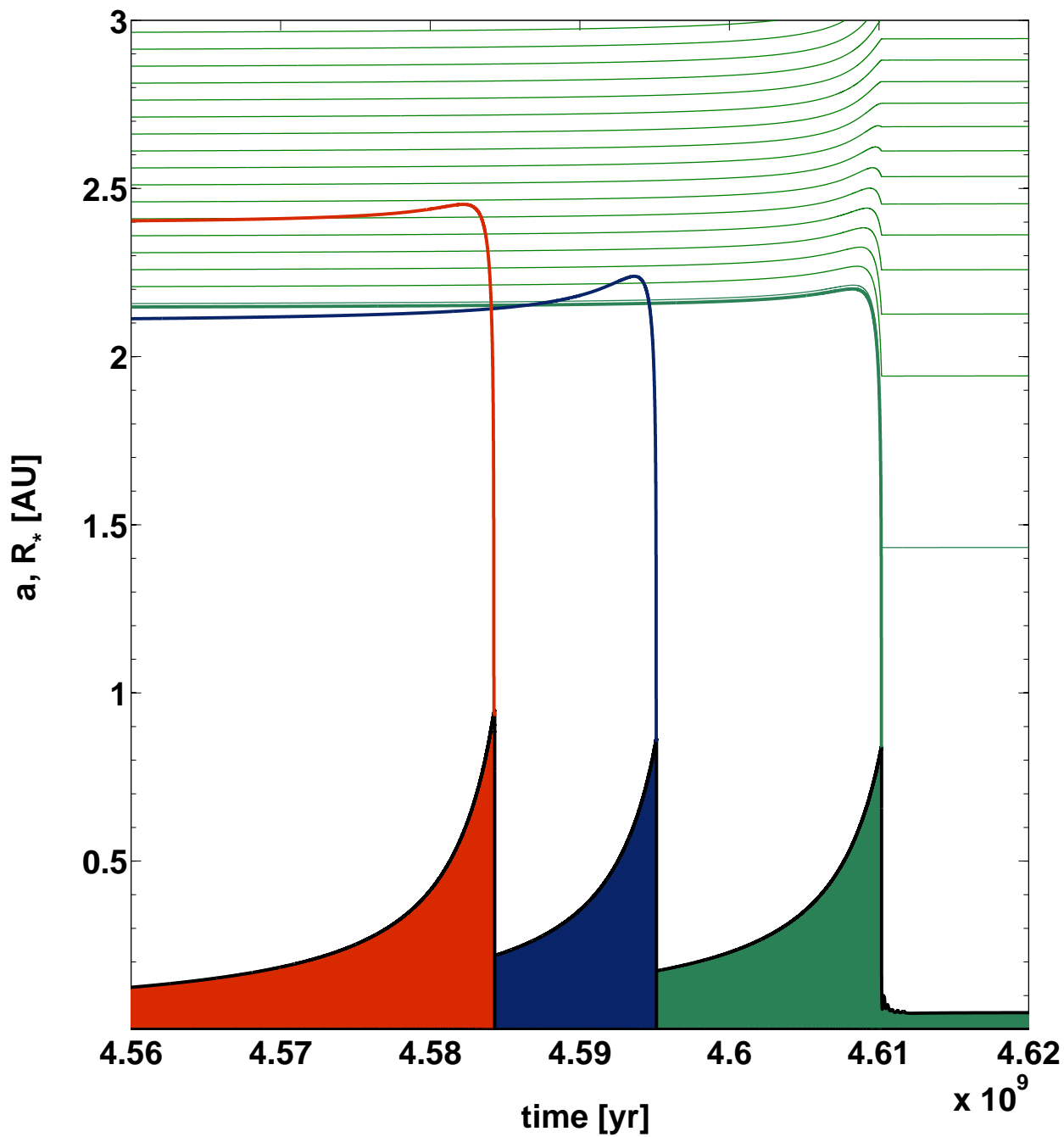}
\caption[ ]{The red ($\eta_R =0.2$), blue ($\eta_R =0.5$), and green (Schr{\"o}der 
\& Cuntz 2005)  areas represent the evolution of the  stellar
radius of a 1.5 \Mso to the RGB tip. The solid lines are the maximum
initial orbital radius for which the planet gets engulfed and are
color coded according to the stellar model. For the Schr{\"o}der 
\& Cuntz (2005) mass-loss prescription model (green) a set of orbits
that avoid engulfment are shown. 
\label{Fig_mases}}
\end{figure}

We tried to capture the main differences in the planet survival limit among the models in
Fig.~\ref{Fig_mases} where we  show  for the 1.5 \Mso
star the different evolution of the
stellar radius for the three mass-loss prescriptions considered (in
red, blue and green for  $\eta_R =0.2$, $\eta_R =0.5$, and
\citealt{Sch05} respectively) and how this leads to different orbital
evolution for a Jupiter mass planet. 
The evolution of the maximum initial orbit that enters the stellar envelope 
is shown for each the mass-loss prescriptions considered. For the \cite{Sch05} model (in green)
we have plotted as well as a set of initial orbits representative of the different
possible outcomes. 

The evolutionary sequences
are indistinguishable in terms of the stellar radius during the subgiant
phase when mass-loss is negligible and the models do not show
significant differences (but not when the star starts the ascent onto the RGB).
Typical values of the radius at the RGB tip are given for the different stellar models in Table~1.

Mass-loss influences planet survival due to three combined
effects: (i) it modifies the stellar radius, (ii) it changes the
orbital angular momentum loss efficiency, and (iii)
it changes the stellar mass and thus its evolutionary timescales. 
When comparing identical mass-loss prescriptions, if
the mass-loss rate is higher (the $\eta_R = 0.5$ case), the pressure on the core is reduced
and off-center helium ignition is delayed. The star reaches smaller radii
and spends more time on the RGB. Lower
mass-loss rates under the same prescription for a given stellar and planet mass, lead to more
distant planets plunging into the stellar surface due to the
combined effects mentioned above. This is why using the lowest mass-loss rate 
simulation, the one using Reimers with
$\eta_R=0.2$, we obtain the largest critical distance for engulfment
(see Fig.~\ref{Fig_mases}). 
The impact of mass-loss on the stellar properties, however, is a highly
non-linear process that depends on how fast the mass-loss
accelerates along the RGB. Note the longer resulting RGB evolutionary timescales obtained for the \cite{Sch05} model. 

The minimum initial orbital distances that a Jupiter mass planet has to have in order to avoid engulfment are given in the last column of Table~1, for the different stellar models used in the calculation. We obtain critical values for the engulfment radius that are always in the upper envelope of those  plotted by \cite{Kun11} in their figures (for stars M$\ge$ 1.8--2\Mso) but are consistent with those obtained for stars M$\in$ [1.5--1.7] \Mso. Given the simplification for the tidal forces these authors made in their paper we have tried to compute the $a_{crit}$ obtained from their Eq. 7 to see if this could be the cause of the difference. We obtain using this equation unrealistic critical distances using the numbers provided by our stellar models, the a$_{crit}$ we get under their prescription are very different from the values we obtain with our full calculation. Generally, we find it hard to reproduce the numbers plotted in their graphs.

In their computation for the critical distance \cite{Kun11} argued that the differences between the critical distance for engulfment obtained by \cite{Vl09} and their models was due to the inclusion of overshooting. In order to check this assertion we have computed 3 models (for M$_\star$=1.5, 1.7 and 1.9 \Mso) using $\eta_R=0.5$ and including overshooting with the same prescription by \cite{Kun11} ($\eta_R=0.5+OS$ in Table.~1) . The critical distances we obtain are similar to the ones obtained in our models without overshooting $\eta_R=0.5$.  We believe that the major differences between ours and the \cite{Kun11} models are a consequence of having transition between degenerate and non-degenerate cores at lower stellar masses, and that cannot be attributable only to the inclusion of overshooting in the models.  

\subsection{Stellar Mass}

\begin{figure}
\epsscale{1.10}
\plotone{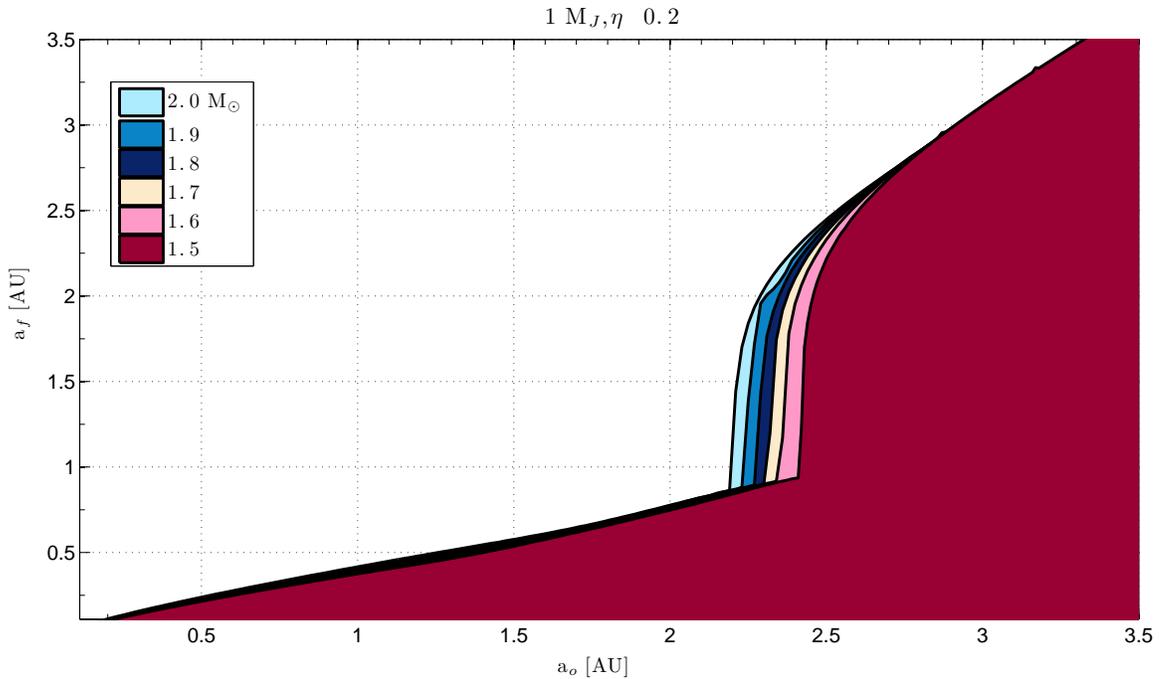}
\caption[ ]{Final ($a_f$) versus initial ($a_o$) orbit reached by a planet of a
  Jupiter mass at the end of the RGB for stellar models with masses
  between 1.5 and 2 \Mso. The mass-loss prescription adopted for the models shown is that
  of Reimers with $\eta_R =0.2$. The different area colors represent
  the different stellar masses as indicated in the legend shown in the
  upper left corner. Planets with final orbits equal or larger than
  the initial ones survive the RGB evolution of the star. Planets with
  initial orbits smaller than that at the inflexion point in the
  curves are engulfed by the stellar envelope. Planets with initial orbits in the
  narrow range between
  the inflexion point and the 1:1 relation experienced tidal decay but
  remain outside the stellar envelope as the latter contracted following He
  core ignition. 
\label{Fig_stmasas}
}
\end{figure}

The stellar mass explicitly enters into the orbital evolution
calculations, affecting the relative strength of the different terms
in the angular 
momentum conservation equations (with the exception of the drag and frictional
forces).  Furthermore, the mass of the star affects both the evolution of the radius 
and the maximum radius reached at the end of the RGB. We have quantified the
variations in the planet survival distance induced by the stellar mass
in Fig.~\ref{Fig_stmasas},  where we show the final orbit reached by the planet ($a_f$)
versus the initial ($a_o$)  one. This figure
is more easily interpreted if compared to  Fig. \ref{Fig1_evo}, 
the linear part of Fig.~\ref{Fig_stmasas} represents the range of initial orbits that will end up at the stellar
surface, the orbits that experience tidal decay but avoid engulfment are those beyond the
inflection point in the figure, and finally those orbits that
experience orbital expansion due to mass-loss are those that satisfy $a_f>a_o$. The different colors
represent the stellar mass (see the
legend at the top left corner of the plot). We have chosen for this
plot stellar models with 
the Reimers's mass-loss prescription and the $\eta_R$=0.2 parameter. 
All the orbits shown in Fig.~\ref{Fig_stmasas} have been integrated for a Jupiter mass
planet. 

Final values of the orbital distance larger than the initial ones
imply that a planet initially 
located at this distance will experience an expansion of
the orbit (note that the planets considered are very close to the star
and unbinding is not expected; e.g. \citealt{Vl07}). That is the case for initial orbital distances $a_o \ge
2.5 \,AU$ for all the masses. Final orbital values smaller than the
initial ones have two possible interpretations: (i) the planet has entered the stellar envelope and
the final orbit simply reflects the value of the stellar radius (where
the calculation is terminated) or (ii) the planet has suffered tidal decay reaching an orbit that avoided the stellar
envelope when He-core ignition took place. These two scenarios are well separated in the plot
by the inflection point in the curves. The main effect of
varying the stellar mass is around the region in the plot where we find orbital decay but the
planet avoids the stellar envelope. Note that for initial orbits
$a_o<2.2 \,AU$ there is no difference in the final orbit induced by
the stellar mass but around $2.2 \,AU \le a_o \le 2.5 \,AU$ increasing the stellar mass
leads to larger final orbital distances which translates into more chances for planet survival. 

A larger stellar mass decreases the influence of the mass-loss term (and its tendency to move the orbit
outward) and decreases the strength of the tidal
forces, effects that 
operate in opposite directions for planet survival (see
Eq. \ref{todo}).  Fig.~\ref{Fig_stmasas}
shows that the stellar mass has a
stronger influence through the tidal term than through the mass-loss
term. This is reflected by the larger distance from the star cleared from planets (smaller color area in Fig.~\ref{Fig_stmasas}) result of the evolution of the stars with lower masses. Note also that we have chosen the
least favorable scenario, the models with the smallest mass-loss
rates, to show this effect. The larger the area shown beyond the
inflection point the better the chances for planet survival.

\subsection{Planet Mass}

\begin{figure}
\epsscale{1.10}
\plotone{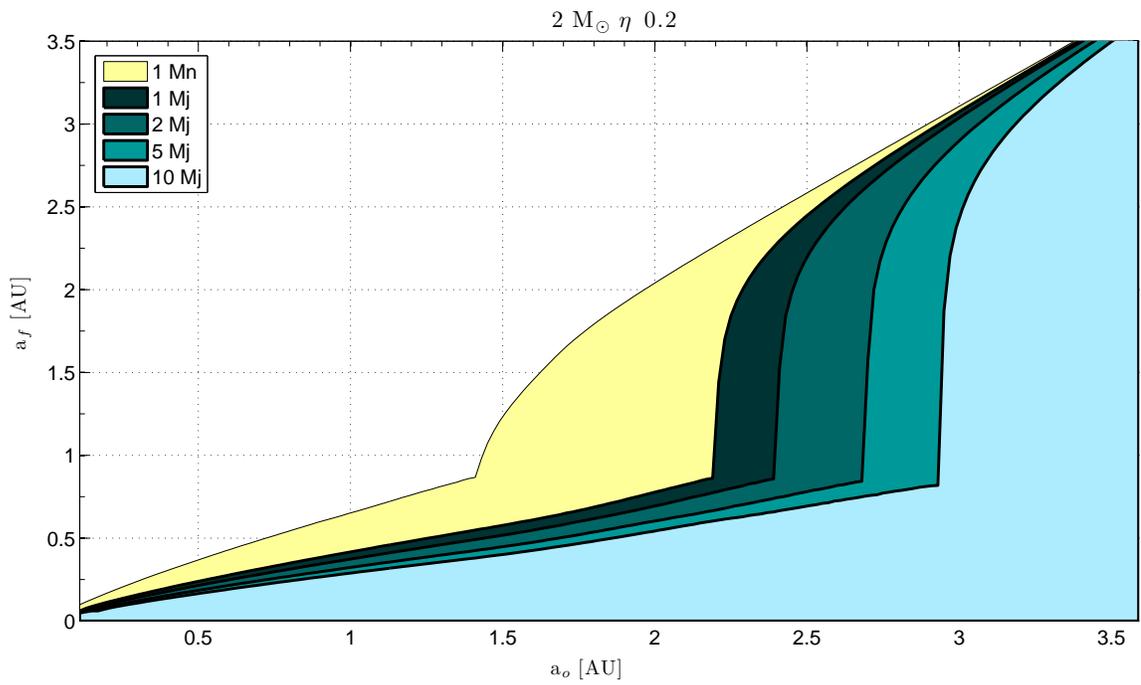}
\caption[ ]{Same as Fig. \ref{Fig_stmasas} but using a star with 2 \Mso,
  Reimer´s mass-loss with $\eta_R =0.2$, and varying the planet mass
  between Neptune mass (1M$_N$ in the legend in the top left corner) and 1, 2, 5 and 10 Jovian
  masses (10 M$_J$ light blue area). 
\label{Fig_plmasas}
}
\end{figure}

The dependency of the of planet survival on the RGB as a function of the planet mass using Neptunian 
and 1, 2, 5, and 10 M$_J$ (Jupiter Mass) planets is shown in
Fig.~\ref{Fig_plmasas}  for a 2 \Mso star with Reimers mass-loss  and
$\eta_R$=0.2.  As already noted in VL09, for a low mass planet to be engulfed, its initial separation
needs to be smaller than that of a more massive planet. This is because at equal
initial separation from
the star, a more massive companion exerts a stronger tidal torque (due
to the dependency on $q$ in Eq.\ref{eq:adotstar}, \ref{tidal}). Note the effect of the 
planet's mass is negligible in the mass-loss term. Overall,  survival is easier for smaller mass planets
around RGB stars since they experience weaker tidal forces.  Fig.~\ref{Fig_plmasas} demonstrates that planet
survival is quite sensitively dependent on the planet's mass. 

Note that the case of massive companions that
can tidally transfer enough angular momentum to the star and 
significantly spin up the primary \citep{Gv14} has not been
considered. In such cases, enhanced mass-loss and deformation of the
primary might take place  as well.  \cite{Nor13} estimated
that this behavior starts to be important for M$_c$/M$_* >$ 0.1, with
M$_c$ the mass of the companion, and excluded this region of
parameter space from their calculations. Note that the largest M$_c$/M$_*$
we use is 6$\times 10^{-3}$, well under the \cite{Nor13} estimated limit.

\begin{figure}
\epsscale{1.10}
\plotone{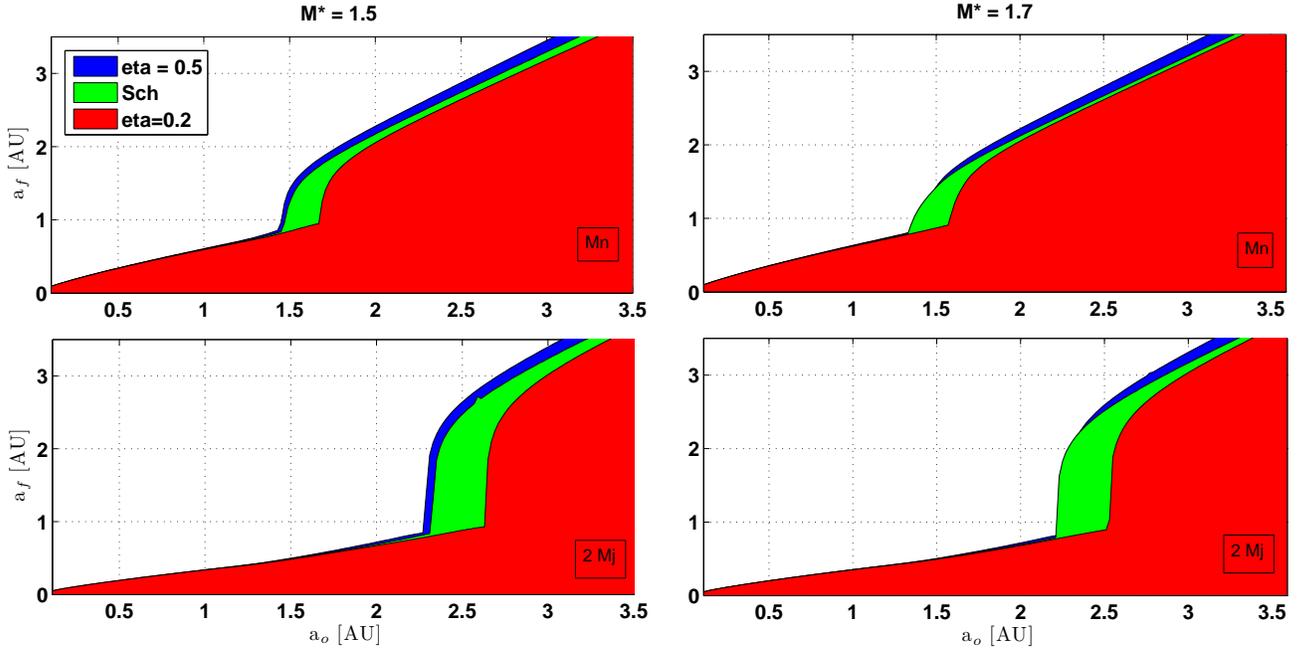}
\caption[ ]{Left column, variation under the mass-loss prescriptions used
  in the stellar evolution calculations of the final ($a_f$) versus initial ($a_o$) orbit of
  a planet orbiting a 1.5 \Mso star.  The top panels are for a planet with Neptune mass and
  the bottom plots are for planets with twice the mass of
  Jupiter. Right column, the same but for a star with 1.7
  \Mso. The legend at the top left corner gives the mass-loss prescriptions used in the calculation
  of the stellar structure. The mass of
  the star is indicated at the top of each column and the planet mass
  at the bottom right corner of each plot. 
\label{Fig_combi}
}
\end{figure}

In Fig.~\ref{Fig_combi}  we show a sample of the computations to 
demonstrate the various effects. The left and right columns show the
effect of varying the stellar mass with different planet masses, 
represented in the top and bottom panels of each column. Within each plot,
the different mass-loss prescriptions used are shown using different
colors. This figure summarizes our findings: i) that planet survival
depends strongly on the planet mass and ii) that the effect
of varying the planet mass
dominates over changes in the mass-loss prescriptions or the stellar
mass in the range considered. 
 
The observational determination of planetary masses through the radial velocity technique 
carries the
uncertainty of an unknown
inclination angle of the orbit, and thus only minimum planet masses are
available. Uncertainties in
stellar masses also contribute to uncertainties in the planet mass
determinations. We show that the largest uncertainties in the outcome
of the process of planet survival during the RGB might be associated
to the uncertainties in the planet masses and not to
uncertainties in the stellar mass determination nor the mass-loss
prescription used in the models.

\subsection{Eccentricity Evolution}
In order to consider planet's eccentricity variations we
integrate equations (\ref{todo}--\ref{gs}). For the eccentricity study we
have chosen two values of the mass of the
planet ($1\mathrm{M_J},1\mathrm{M_N}$),  two stellar masses 
($1.5,2\mathrm{M}_\odot$), and $\eta_R$=0.5. Note that mass-loss 
has little effect on the planet's eccentricity, since the mass-loss rates are rather small
and the orbits considered here are relatively close to the star (e.g. \citealt{Ver11}). 
The planet's spin rate was in all cases initially 100 radians/yr but we verified that the exact initial value is
unimportant. 

The planet's radius was fixed at 1 Jupiter radius
($1\mathrm{R_J}$) or 1$\mathrm{M_N}$ for the Neptune-mass planet; we did
not consider the planet's radius expansion due to tidal energy being
deposited in the planet's interior. The planetary tidal strength was
set to $Q_\mathrm{pl}^\prime=10^5$ and the stellar properties were
evolved using the models as described in  \S2.1 from the pre-MS. As before, the
star in our models does not spin.

As noted above, for the study of the eccentricity evolution we kept the planet's
radius, $R_\mathrm{p}$, fixed in
Eqs. \ref{eq:edotpl} to \ref{eq:odotpl}. In real systems,
heating of a planet's interior by tidal forces can cause the planet to expand, further
strengthening tidal forces and potentially even leading to a runaway process. This
effect, important to consider in detailed studies of the evolution of
MS close-in planets would have little importance for the overall outcome of
the systems considered in this paper: if they are close enough
for substantial tidal heating they would be engulfed very early by the expansion of the star. 
Furthermore, for simplicity, we have removed from Eq. \ref{todo} the second term
involving the frictional and gravitational drag forces since we have seen in the previous part of the study they have
a negligible effect in reducing the semi-major axis over the RGB lifetime.

Some comments on the qualitative behavior of the effects of stellar
and planetary tides are in order. The evolution the eccentricity
during the  MS, when the stellar radius is small, 
is dominated by planetary tides (e.g. \citealt{Matsu10}). During that
period, the planet is rapidly brought into a state of pseudosynchronisation, where 
\begin{equation}
\Omega_\mathrm{pl}=\frac{g_2}{g_5\left(1-e^2\right)^{3/2}}n.
\end{equation}
Once this is attained, the planet's eccentricity decays while
preserving the orbit's angular momentum, following a trajectory in the 
$a-e$ space 
\begin{equation}\label{eq:a-e-planet}
e=\sqrt{1-a_f/a}.
\end{equation}
Here, the eccentricity decays to zero as the semi-major axis decays to a finite value, $a_f$.
In contrast, when the stellar radius is much larger, on the RGB,
stellar tides dominate the evolution. Neglecting frequency
dependencies for simplicity, the planet's orbit follows a trajectory, 
\begin{equation}\label{eq:a-e-star}
e=e_0\left(\frac{a}{a_f}\right)^{9/2}.
\end{equation}
where $e_0$ is the initial eccentricity. Now the eccentricity decays to zero as the semi-major axis decays to
zero. This qualitatively different behavior of the $a- e$ trajectories
under stellar and planetary tides has important implications for the
shape of the $a-e$ envelope, as discussed later.

\begin{figure}
\plotone{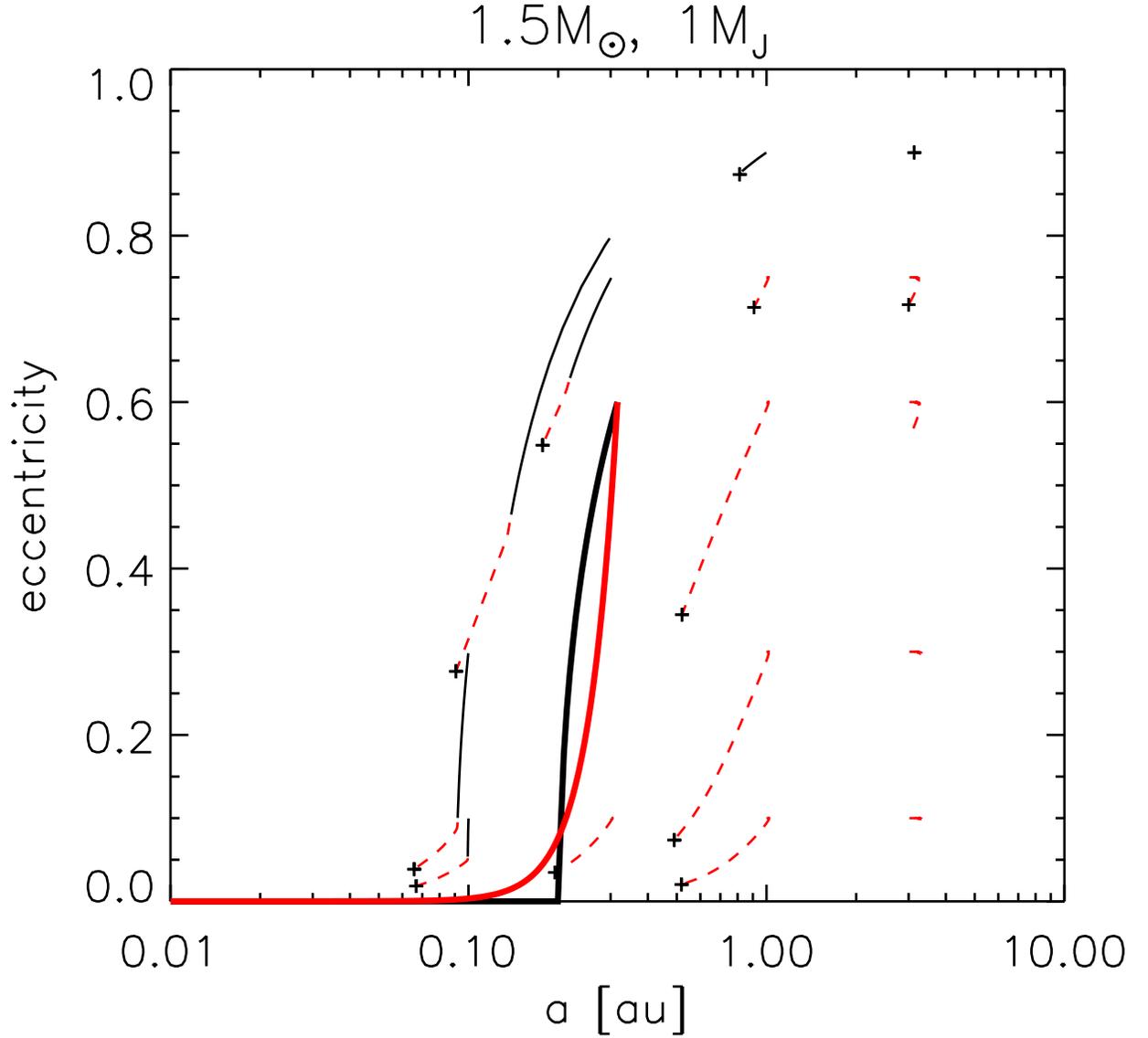}
\caption[ ]{Sample evolutionary tracks in $a-e$ space of planets of
  mass $1\mathrm{\,M_J}$  orbiting a $1.5 \mathrm{\,M}_\odot$
  star. Linestyles show different evolutionary phases: solid
  black line MS and dashed line the RGB. Crosses indicate where
  planets are engulfed by the star. The solid and red thick lines show
the analytical approximation for the evolution on the MS and RGB
respectively (Eqs. 16 and 17).}
  \label{fig:tracks}
\end{figure}

In Figure~\ref{fig:tracks}, we show samples of the evolution
of a  $1\mathrm{M_J}$ planet on the $a-e$ plane. Evolution during the
stellar MS is shown as black solid lines and on the RGB as
red-dashed lines. On the MS, the planet's path is determined by
planetary tides and it follows Equation~\ref{eq:a-e-planet},  which
brings the eccentricity down steeply while typically causing a modest
decay in semi-major axis. An illustration of this path is shown in Fig.~\ref{fig:tracks} as a black thick
line. On the RGB, however, stellar tides dominate
and the track follows a shallower path as $a$ and $e$ decay at similar
rates (see red solid thick line). The difference in the tracks is most clearly seen in the planets
starting at 0.1 \, AU in Fig.~\ref{fig:tracks}. Very distant planets do
not experience tides and their orbits expand due to mass-loss. Planets at
intermediate distances at around 1\,AU execute a hook, as they first
see their orbits expand due to mass loss but then decay tidally as the
stellar radius grows. Most of the planets plotted in Fig.~\ref{fig:tracks} are eventually
engulfed on the RGB. 
 
\begin{figure}
  \plotone{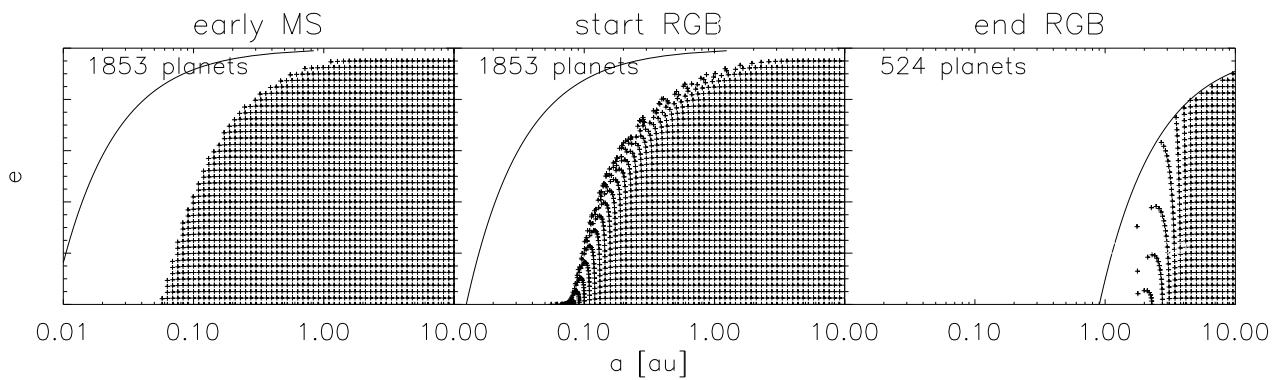}
  \caption{Populations of $1\mathrm{M_J}$ planets orbiting
    $1.5\mathrm{M}_\odot$ stars at three points in the star's life:
    early MS (left) $\approx$ 20 Myr, beginning of RGB (center), and end of RGB
    (right). The solid line marks the locus where the planets'
    pericenters  lie on the stellar surface. The number of planets is
    shown in each panel}
  \label{fig:pops}
\end{figure}

Populations of planets in the $a- e$ plane are shown in Figure~\ref{fig:pops}. 
To make this plot we have filled the region $a\in [0.01, 10]$ \,AU and
$e\in[0, 0.95]$  (with $e$ increasing in steps of 0.05) with 2964 planets and let the star and the orbit evolve. 
Planets with pericenters inside the stellar envelope at any time are then removed.
We show the populations at meaningful times in the star's life: very
early in the MS (at $\approx 20 Myr$), to show the set of initial conditions being integrated, i.e. those
that did not have their pericenters inside the stellar envelope during
pre-MS contraction; at the start of the RGB (defined by $R_\star=
3.4R_\odot$), showing the extent of eccentricity decay during the MS,
largely due to planetary tides; and at the end of the RGB, showing the
clearing effects of stellar tides and the expanded stellar
envelope. We included the star's pre-MS contraction phase in the
calculation which  limited the initial pericenters of our planet
population (note we run models for the $a-e$ parameter space described above). 
By following the pre-MS and MS evolution the goal was not to study
those phases in detail but rather to get some reasonably realistic initial
conditions for the post-MS phase.  Note that, while the envelope expands to only $\sim 1$ AU, the
region out to 2-3 AU shows significant depletion of the planetary
population.  
 
\begin{figure}
  \plotone{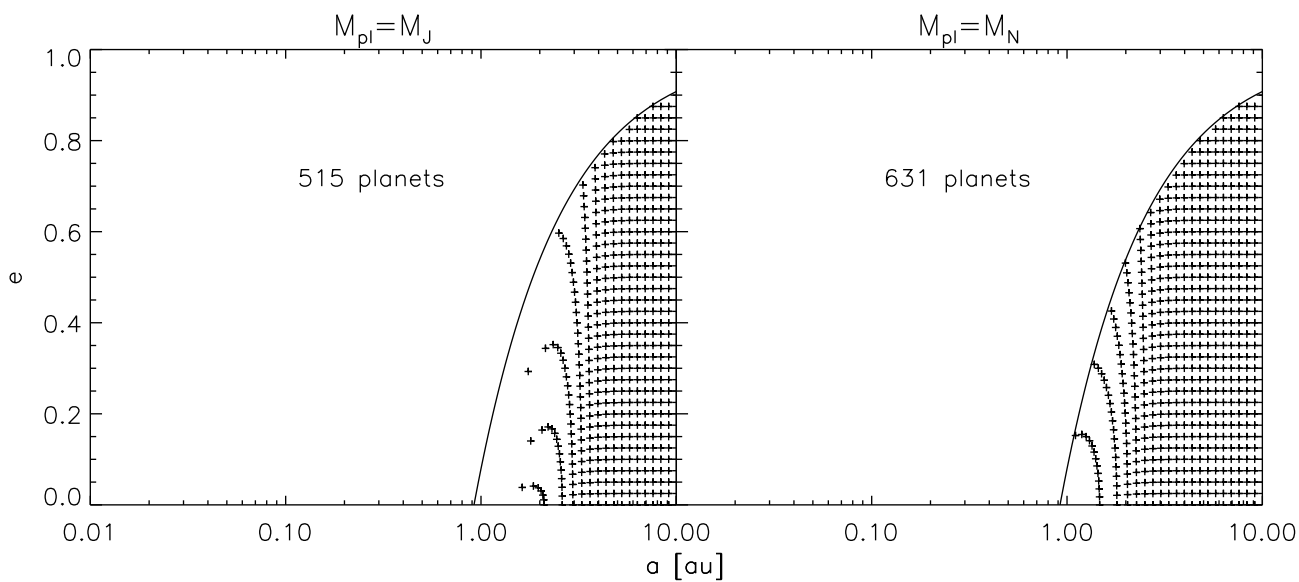}
  \caption{The effects of changing the planet mass on the planet
    population surviving the RGB. Left to right: $2 \mathrm{M}_\odot$
    and $1\mathrm{M_J}$; $2 \mathrm{M}_\odot$ and $1\mathrm{M_N}$.} 
  \label{fig:morepops}
\end{figure}

Figure~\ref{fig:morepops} shows the effects of varying the 
planetary mass on the distribution of orbital properties at the end of the RGB. 
As we have previously discussed we see a larger initial region of
orbital distances resulting in
planets getting accreted onto the stellar surface increasing as the
planet mass increases. In general, we 
find that on the RGB, very eccentric and distant planets (a $\gtrsim$ 3
AU, e $\gtrsim$ 0.8) do not see
much eccentricity decay,  and their engulfment is basically just
determined by their pericentre location and the maximum stellar
radius. 

\begin{figure}
  \includegraphics[width=.9\textwidth]{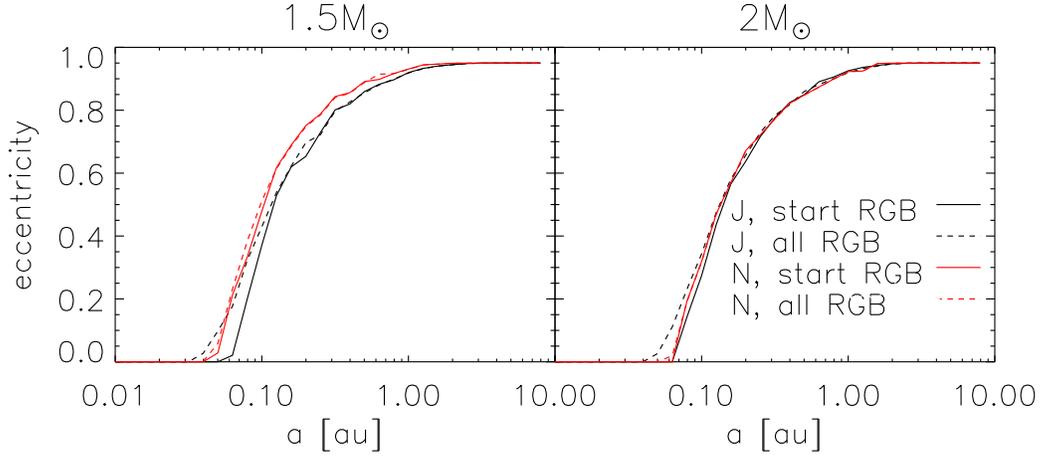}
  \caption{Envelopes of planet populations around
    $1.5\mathrm{M}_\odot$ and $2 \mathrm{M}_\odot$ stars (left and
    right panels respectively). Envelopes of the population at the
    beginning of the RGB are shown as solid lines, while envelopes
    over the \emph{whole} RGB population are shown as dashed
    lines. Stellar tides cause some Jovian planets to decay through a
    region of moderate eccentricity and small semi-major axis that was
    depleted during MS evolution by planetary tides.} 
  \label{fig:envelopesdual}
\end{figure}

Finally, in Figs.~\ref{fig:envelopesdual} we show how stellar tidal forces on
the early RGB can create a transient population of moderately
eccentric close-in Jovian planets. In Fig.~\ref{fig:envelopesdual} the envelope for \textit{'start RGB'} is defined such that there are no planets
lying above the envelope  at the initial RGB output timestep.
The envelope for \textit{'all RGB'} is defined such that there are no planets
lying above the envelope at any RGB output timestep, i.e. at no point
on the RGB is a planet found above the envelope. For
every point below the envelope, there exists at least one time, $t$, such
that a planet is found there.

Eccentric planets with
$a\lesssim0.05$ AU are depleted on the MS due to the action of
planetary tides. However, as discussed above, planetary and stellar
tidal forces cause planets to follow different tracks in $a-e$ plane. Once the stellar radius begins to increase on the RGB, stellar
tidal forces begin to dominate over planetary tidal forces, and
the planets now follow tracks through the region depleted on the MS, with
$a\sim0.05$ AU and $e\sim0.2$. Planets following these tracks, however,
are rapidly engulfed by the expanding stellar envelope. Neptune-mass
planets do not exhibit this behavior, since the stellar tides they
excite are much weaker. 

\section{DISCUSSION}

\begin{figure}
\epsscale{1.10}
\plotone{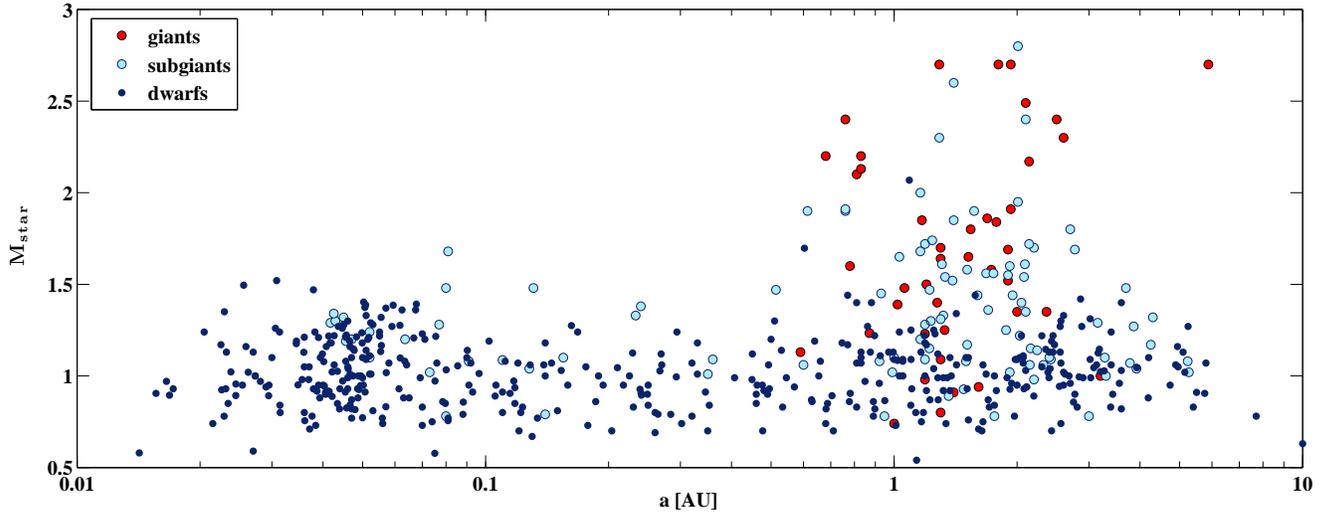}
\caption[ ]{The observed orbital distance versus the stellar mass of all
 the confirmed planets detected through the radial velocity method
  taken from the Exoplanet encyclopedia (exoplanet.eu,
  exoplanets.org). The different colors represent the evolutionary
  status of the star as determined by the published luminosity class. 
\label{fig_obsmstar}}
\end{figure}

The semi-major axis distribution of planets discovered with RV measurements radial
velocity surveys as of the 1st of September 2013\footnote{Data 
from the exoplanet.eu, exoplanets.org \citep{exo11,exorbit11}} is plotted versus the
stellar mass in Fig.\ref{fig_obsmstar}.  The points have been color coded according
to the evolutionary status of the star, with dark blue, light blue and
red representing main sequence, subgiants and red giant
stars, respectively.

The planet population orbiting massive host stars (meaning M
$\ge 1.5$ \Mso) has been claimed to be distinct with respect to the minimum orbital distances at
which planets are found \citep[see e.g.,][]{Joh10, Sat08, Bow10}. Indeed
no planet around a giant has been found so far with $ a < $ 0.54 \, AU
\footnote{The planets
orbiting BD+15 2940 at 0.54 AU \citep{Now13},  HIP 63242 at 0.57 AU
\citep{Joh13} hold the record.} (from radial
velocity measurements). This fact (taken alone) has been used
to suggest a relation between
planet formation and  the stellar mass
\citep{Cur09}. However, as we see in Fig. \ref{fig_obsmstar},
planets orbiting subgiant stars populate the region $a \le$ 0.54 \,AU
and can be found as close as 0.08 \, AU orbiting the subgiant star
HD 102956 \citep{Joh10}. Furthermore,  recently transit surveys have discovered  hot jupiters
around Main Sequence A-F stars that remained elusive for radial velocity surveys
(e.g. HAT-P-49 \citealt{Bier14}; WTS-1b \citealt{Cap12}; Kepler-14b \citealt{Buch11}; WASP-33 \citealt{Coca10}; KELT-3b
\citealt{Pep13}; COGLE-TR-L9 \citealt{Sne09}). These are all mostly fairly
low mass stars (1.5, 1.2, 1.5, 1.5, 1.3, and 1.5 \Mso). The best
example so far of a close-in planet
around a high-mass star is HD 102956b orbiting a  1.7 \Mso star \citep{Joh10}.  A transiting planet, Kepler-91b
has been found at 0.072 \,AU from the star \citep{Lillo13}  as well, holding the record for the
innermost planet found around a RGB star.  Although note that the
stellar mass is 1.3 \Mso, a relatively low value compared to most  RGB planet host stars, and recent
claims have been made of a false positive for this system\citep{Sli,Est13}. 

\begin{figure}
\plotone{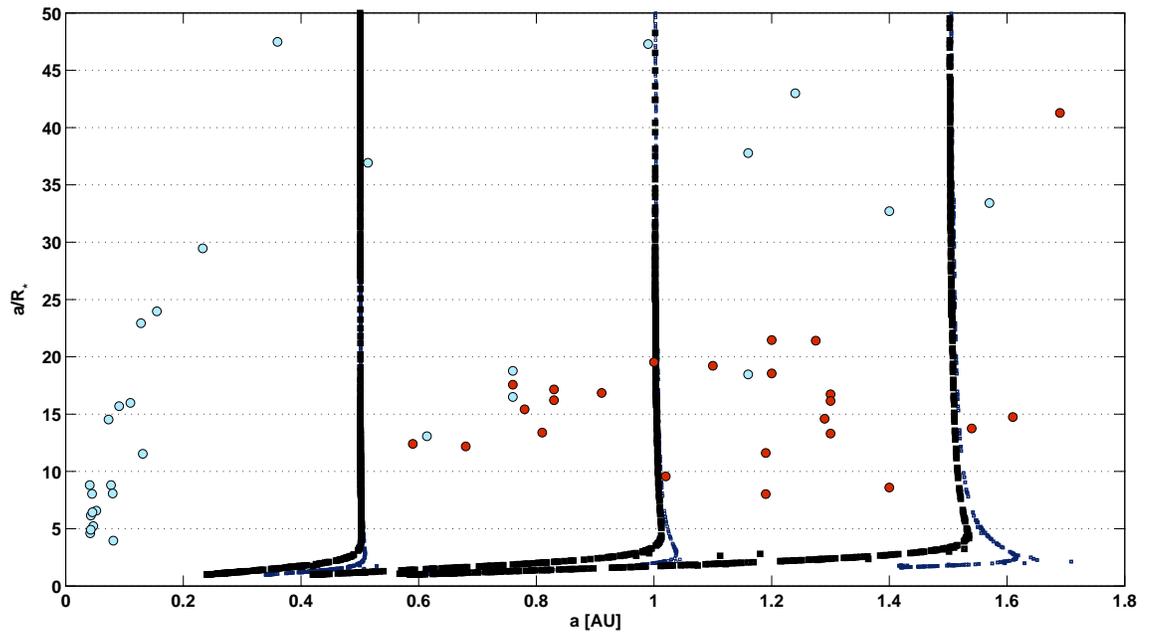}
\caption[ ]{Ratio of the orbital distance to the stellar radius versus the orbital
  distance of the observed population of planets orbiting subgiant
  stars (light blue points) and giant stars (red points). The
  evolution of the orbital to stellar ratio is shown for 3 initial
  orbital distance (0.5, 1 and 1.5 \,AU) and two stellar masses 1.5
  \Mso (in dark blue) and 2 \Mso  (in black) and for a jupiter mass planet.
\label{ar}}
\end{figure}

If close-in planets are present in the
MS stage orbiting A-F stars but not found from radial velocity searches around evolved
stars, this suggests that it is the evolution of the star, and not
its mass, that play a role in removing planets from close orbits. In fact,
as we have shown here, close-in planets enter the stellar envelope as the
star leaves the main sequence and evolves onto the RGB (see as well \citealt{Vl09,Kun11}). 
As the star evolves, it removes 
planets from a region that extends far beyond the stellar radius to
the entire region of tidal influence ($a/R_*\approx$ 2-3; see
Fig.~\ref{ar} for a Jupiter mass planet). The star first, during the  sub-giant phase, clears
out the very close-in planets present during the MS evolution and then
proceeds to clear out a larger region as the stellar radius increases
when it ascends the RGB. The tidal influence region is where
planets are expected to be depleted from as the star evolves. Note in
Fig.~\ref{ar} that most observed planets are safe from being engulfed
since they are located far beyond this region. Theoretical predictions agree with the
observations since there is just one planet detection with $a/R_*\le$ 2.5 (the closest one  Kepler-91b
has  $a/R_*\approx$ 2.45 \citealt{Lillo13}).

A feature that is harder to understand in Fig.~\ref{fig_obsmstar} is the
absence of planets in the upper middle region in the plot (no planets
found around  subgiant stars with $0.08 \le a_o \le 0.5  \, AU$) and M $\ge$ 1.5 \Mso. Planets are found close to massive 
subgiant stars, and around them beyond 0.5 \, AU but no subgiant with M $\ge$ 1.5 \Mso has been found 
with planets in that narrow range of orbits. Planets are not expected to be cleared out that far by tides during the evolution of the star along the subgiant phase. These planets are expected to be engulfed by the star later on,
during the RGB evolution. So the question is, why are there planets  found in the  $0.08 \le a_o \le 0.5  \, AU$ orbital range around less massive subgiants but not around subgiants with M $\ge$ 1.5 \Mso? 

In this paper we have explored a host mass
range between 1.5 and 2 \Mso for planet survival. Within this mass range we find that during the subgiant phase
planet survival is not substantially affected either by the stellar mass, or the
mass-loss prescription adopted in the calculation within the reasonable assumptions adopted for the stellar models.  The main parameter affecting planet survival during the subgiant phase is the planet mass. More massive planets are more likely to plunge earlier into the stellar envelope. More massive planets are formed
preferentially in more massive protoplanetary disks and thus those planets are expected around more massive stars.
If more massive planets are present around subgiants with M $\ge$ 1.5 \Mso then they could be brought into the stellar envelope early in the evolution if they are initially located at relatively close distance from the star and this could explain why they are not found  in the  $0.08 \le a_o \le 0.5  \, AU$ orbital range. Less massive planets are harder to detect around this systems.  There is an observational bias
associated with the large stellar jitter in evolved hosts that prevents the detection of low mass planets around evolved stars. 
In fact more massive planets are found on average orbiting evolved
stars (which, in principle, are more massive than the average main sequence planet
hosts). The lack of observed planets around subgiants with M $\ge$ 1.5 \Mso in the  $0.08 \le a_o \le 0.5  \, AU$ 
orbital range is hard to understand otherwise from an theoretical perspective.

The RGB evolution is short compared to the MS lifetime, implying that the
RGB region in the HR diagram is
populated by stars of similar mass (for a given initial chemical
composition and age). Masses for RGB (planet hosting) stars are computed from
spectroscopic values of Teff and metallicities that together with
visual magnitude estimates are used to interpolate in theoretical stellar
evolution grids. The uncertainties involved in the
measured parameters are therefore relatively large. Furthermore, the tracks for the
different masses in the HR diagram are packed together and as a
consequence, estimates of masses for RGB stars carry large uncertainties.  
Uncertainties in mass are for the most part $\le$ 0.1 ( $\le$ 10\% see e.g. \citealt{Mal13,Mor13,Zie12}) 
with some objects having larger  uncertainties (up to 0.3) mostly caused from the lack of available 
Hipparcos parallaxes \cite{van Leeuwen (2007)}.  The stellar radius have the same dominant source 
for the uncertainty as the stellar mass and have values of  $\approx$ 6--15 \%.
We have shown that mass differences within the studies range do not have a
strong influence on planet survival along the RGB and thus mass observed mass uncertainties 
should not have a strong influence in the outcome of the distribution of observed planets. 

It has been argued that the planet survival limit is rather sensitive
to the host's star mass \citep{Kun11}. While this is true when the stellar
mass crosses from  low to intermediate mass  (where the stellar core
conditions change from degenerate to not degenerate),  it is not
true when low mass stars are involved. Along these lines it is important to remember 
that while the RGB T$_{eff}$ at a given luminosity is only slightly sensitive to the 
value of the stellar mass, it is affected by the initial value of the metallicity $Z$. Increasing
$Z$ moves the transition between low and
intermediate mass stars towards higher masses. Higher mass stars would
ascend the RGB faster and grow smaller in radius (note that this
transition mass is set al 1.4 \Mso in
\cite{Ll13}  where he argues for lower mass planet hosting evolved
stars  than claimed in the literature). Our limit for He-flash stars is
at 2.5 \Mso for [Fe/H]= 0.19. It
is the mass limit between degenerate and non-degenerate cores that strongly influences the planet survival
limit during the RGB.  Close-in planets thus would have more chances to survive
orbiting metal poor stars due to this effect. Note the mass limit to
build a degenerate core during the RGB is 1.5 \Mso for Z=0.01 
$Z_{\odot}$ \citep{Swe90}. 

We have shown that for a given mass-loss prescription, increasing the
stellar mass puts more planets into the safe zone limit. This is
because the stellar mass has a stronger influence (through the tidal term) 
than the mass-loss term. 

\begin{figure}
\plotone{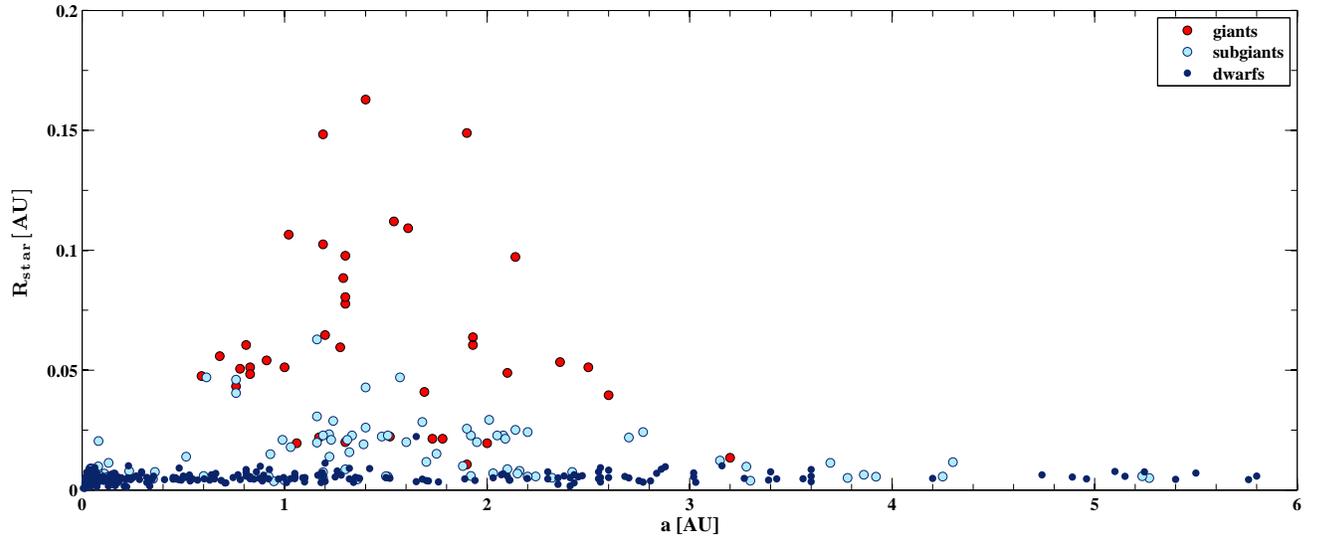}
\caption[ ]{Same as Fig. \ref{fig_obsmstar} but here the observed
  orbital distance is plotted against the stellar radius.
\label{fig_obsradio}}
\end{figure}

In Fig. \ref{fig_obsradio}  we show the distribution of stellar radi
as a function of orbital  separation for planets found in radial velocity surveys. The
planets clearly occupy different regions according to the
evolutionary status of the star. In this plot, the striking feature
is the lack of planetary systems found orbiting giant stars at large
distances, beyond 3 \,AU. There is nothing in principle that could cause this region to be
depleted of planets unless stellar jitter is preventing the
detection at large distances or planet searches around giant stars
have not been carried long enough to recover this planet population present, on the other
hand around MS stars. We find that all the observed systems are
within the $a/R_*$ safe zone from tidal interaction we find from the
calculation of the orbital evolution.

\begin{figure}
\plotone{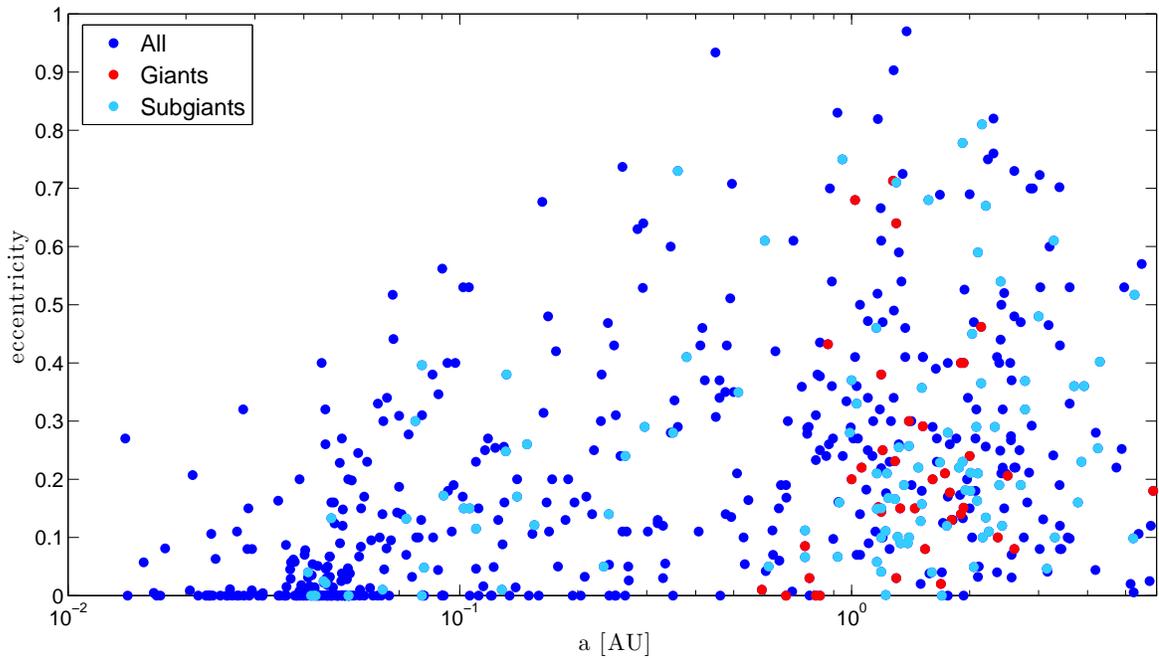}
\caption[ ]{Same as Fig. \ref{fig_obsmstar} but here the observed
  planet eccentricity is plotted versus orbital distance.
\label{fig_ae}}
\end{figure}

Regarding the eccentricities of planet orbits, the observed
distribution is shown
in Fig. \ref{fig_ae}. According to theoretical expectations giant stars
being systems dominated by stellar tides tend to have  $a$ and $e$ decaying at
similar rates. It is expected therefore that planets experiencing tidal
eccentricity decay would as well experience orbital decay and thus
would not be observed. Subgiants and MS stars on the other hand, could
have experienced eccentricity decay but not substantial orbital decay,
explaining the upper left corner clearance zone in the plot.

In summary, our calculations show that rapid tidal orbital decay
occurs when $a/R_*<$3. There are only 3 planets found orbiting RGB
stars with $a/R_*<$10, and none has been found with  $a/R_*<$8. The region $\in$[0.08, 0.5]\,AU 
around subgiants is too wide to have been depleted of planets just by tidal
effects, so the depletion must have other origins and cannot be
primordial since this planets have been found recently orbiting A-F MS
stars using imaging techniques. A mass dependent mechanism must be at
play that  can only act in one way by building more massive planets
are around more massive hosts that due to tidal forces are brought
into the stellar envelope from larger initial orbital distances. The
only way that makes sense is if the RGB stars are indeed higher mass, distinct from the MS
planet hosts. If the RGB stars are low mass, as \cite{Ll13} and \cite{Sw13} 
argue, then there is no sensible way of accounting for the lack of
close-in RGB planets since the stellar radius is too small to cause
depletion by tidal forces (unless the stellar radii are systematically
underestimated, by a factor of 3 or more).  

\section{CONCLUSIONS}

In this work we have quantified the influence of different parameters
on the survival of planets orbiting subgiant and red giant
stars. We have explored a mass
range for the hosts between 1.5 and 2 \Mso and find that during the subgiant phase 
planet survival is not substantially affected either by the stellar
mass, or the adopted
mass-loss prescription.  The main parameter
affecting planet survival during the subgiant phase is the planet
mass, with more massive planets being more likely to plunge earlier into the
stellar envelope.  

We find that even though the observed
uncertainties in the determination of the stellar mass or mass-loss
rates are quite large, those do not have much
influence on the critical orbit beyond which a planet will survive 
RGB evolution. 
Since tidal torques drop as a high negative power of the ratio of
stellar radius to orbital separation, companions that escape engulfment experience
essentially no tidal interactions if they are located beyond a
certain initial orbit. Planets located at $a/R_*\approx$ 2-3 are in
jeopardy as soon as the star leaves the MS since that is where the tidal force
starts to dominate the orbital evolution.

Eccentric planets with
$a\lesssim0.05$ \, AU are depleted on the MS due to the action of
planetary tides.  However, once the stellar radius begins to increase on the RGB, stellar
tidal forces begin to dominate over planetary tidal forces, and
planets follow tracks through the region depleted on the MS, with
$a\sim0.05$ \, AU and $e\sim0.2$. Planets following these tracks, however,
are rapidly swallowed by the expanding stellar envelope. Also
important is that we find that during the RGB very eccentric and
distant planets do not experience much eccentricity decay,  and that
planet engulfment is basically just determined by  pericentre location and the maximum stellar
radius.

\acknowledgments
E.V. and A. J. M.'s work was supported by the Spanish Ministerio de Ciencia e
Innovaci\'on (MICINN), Plan Nacional de Astronom\'{\i}a y Astrof\'{\i}sica, under
grant AYA2010-20630 and by the Marie Curie program under grant
FP7-People-RG268111.  LS is a FNRS research associate.
We thank Amy Bonsor for useful discussions

\clearpage

\begin{tabular}{lclllllclc}
\hline
\hline
$M_\mathrm{ini}$ & wind & $M_\mathrm{1DUP}$ & $M_\mathrm{tip}$ &  $R_\mathrm{tip}$ &  $L_\mathrm{tip}$ &  $M_\mathrm{tip}^\mathrm{core}$ & $\dot M_\mathrm{tip}$ & $t_\mathrm{RGB}$ &$a_c$, $M_p=M_J$\\

[$M_\odot$] &  & [$M_\odot$] & [$M_\odot$] & [$R_\odot$] & [$L_\odot$] & [$M_\odot$] &
[$M_\odot$/yr] & [Myr] & [AU] \\
\hline
1.5 & $\eta_R=0.2$  & 1.497 & 1.410 & 216.2 & 3015.4 & 0.489 & 3.469$\times10^{-8}$ & 203.9 & 2.43  \\
1.5 & $\eta_R=0.5$  & 1.494 & 1.304 & 196.4 & 2560.4 & 0.472 & 7.251$\times10^{-8}$ & 203.6 &   2.11\\
1.5 & $\eta_R=0.5+$OS  & 1.494 & 1.272 & 221.9 & 2913.2 & 0.485 & 9.465$\times10^{-8}$  & 197.5 &2.10\\
1.5 &            Sc  & 1.490 & 1.369 & 190.3 & 2547.7 & 0.471 & 6.834$\times10^{-8}$ & 206.3 &2.14\\
1.6 & $\eta_R=0.2$ & 1.597 & 1.516 & 209.2 & 3027.7 & 0.490 & 3.152$\times10^{-8}$ & 185.1 &2.38\\
1.6 & $\eta_R=0.5$ & 1.594 & 1.393 & 209.4 & 2871.6 & 0.483 & 8.105$\times10^{-8}$ & 186.6 &2.20 \\
1.6 &           Sc  & 1.589 & 1.448 & 209.4 & 2960.1 & 0.486 & 8.924$\times10^{-8}$ & 190.1 &2.04\\
1.7 & $\eta_R=0.2$ & 1.697 & 1.620 & 204.8 & 3072.7 & 0.491 & 2.946$\times10^{-8}$ & 190.2&2.34 \\
1.7 & $\eta_R=0.5$ & 1.694 & 1.519 & 191.2 & 2705.1 & 0.477 & 6.458$\times10^{-8}$ & 190.3 &2.09\\
1.7& $\eta_R=0.5+$OS  & 1.694 & 1.509 & 201.8 & 2888.1 & 0.484 & 7.300$\times10^{-8}$& 154.2 &2.07\\
1.7 &            Sc & 1.688 & 1.590 & 176.8 & 2542.8 & 0.470 & 4.904$\times10^{-8}$ & 176.7 &2.03\\
1.8 & $\eta_R=0.2$ & 1.797 & 1.724 & 202.2 & 3128.0 & 0.493 & 2.791$\times10^{-8}$ & 161.5 &2.29\\
1.8 & $\eta_R=0.5$ & 1.794 & 1.638 & 177.1 & 2569.4 & 0.472 & 5.300$\times10^{-8}$ & 162.5 &1.99\\
1.8 &           Sc  & 1.787 & 1.686 & 181.9 & 2716.6 & 0.477 & 5.108$\times10^{-8}$ & 183.9 &2.09\\
1.9 & $\eta_R=0.2$ & 1.898 & 1.827 & 198.2 & 3155.4 & 0.493 & 2.613$\times10^{-8}$ & 149.9 &2.25\\
1.9 & $\eta_R=0.5$ & 1.894 & 1.727 & 190.1 & 2898.6 & 0.484 & 6.082$\times10^{-8}$ & 149.3 &2.09\\
1.9 &  $\eta_R=0.5+$OS & 1.894 & 1.732 & 191.0 & 2934.7 & 0.486 & 6.170$\times10^{-8}$ & 86.9 &2.09\\
1.9 &           Sc  & 1.885 & 1.786 & 184.2 & 2847.2 & 0.481 & 5.091$\times10^{-8}$ & 146.1 &2.09\\
2.0 & $\eta_R=0.2$ & 1.997 & 1.930 & 193.9 & 3166.5 & 0.494 & 2.434$\times10^{-8}$ & 148.7 &2.21\\
2.0 & $\eta_R=0.5$  & 1.995 & 1.821 & 198.6 & 3155.7 & 0.494 & 6.560$\times10^{-8}$ & 145.9 &2.15\\
2.0 &            Sc  & 1.984 & 1.874 & 196.0 & 3173.7 & 0.494 & 6.032$\times10^{-8}$ & 160.2 &2.19\\
\hline
\hline
\label{tablita}
\end{tabular}


\clearpage

\newpage

\newpage

\newpage

\newpage

\newpage

\newpage

\newpage

\newpage

\newpage

\newpage

\end{document}